%% file: main.tex
\documentclass{article} 
\usepackage{iclr2027_conference,times}

\input{math_commands.tex}

\usepackage{hyperref}
\usepackage{url}
\usepackage{graphicx}
\usepackage{booktabs}
\usepackage{physics}
\usepackage{amsmath}
\usepackage{amssymb}
\usepackage{wrapfig}
\usepackage{algorithm}
\usepackage[noend]{algpseudocode}  
\algrenewcommand\algorithmiccomment[1]{\hfill$\triangleright$ #1} 

\renewcommand{\eqref}[1]{Eq. (\ref{#1})}

\title{Neural Boltzmann Equations}

\author{Jonas Spinner \\
Institute for Particle Physics Phenomenology\\
Durham University, Durham, UK \\
\texttt{jonas.spinner@durham.ac.uk} \\
\And
Jack D. Shergold \\
Department of Mathematical Sciences \\
University of Liverpool, Liverpool, UK \\
\texttt{shergold@liverpool.ac.uk} \\
}

\iclrfinalcopy 
\begin{document}

\maketitle

\begin{abstract}
The dynamics of particles in the early universe are described by Boltzmann equations, which involve high-dimensional phase-space integrals. Classical approaches use quadrature integration and evolve the system on a fixed momentum grid, which scales poorly to complicated systems and parameter scans, severely limiting the complexity of processes that can be studied. We introduce Neural Boltzmann Equations (NBEs), which combine three coupled concepts to overcome these limitations. First, particle properties are encoded in physics-inspired neural distribution functions, with parameters that can be predicted using neural networks, enabling efficient parameter scans. Second, phase-space integrals are evaluated with Monte Carlo, using importance sampling tools from collider physics. Third, we use the natural gradient method to evolve the system. After demonstrating the individual benefits of NBEs, we use the framework to perform the most numerically precise calculation to date of the effective number of relativistic neutrino degrees of freedom, $N_\mathrm{eff}$, in the Standard Model.
\end{abstract}

\section{Introduction}

Cosmology is entering a precision era, with ongoing and upcoming experiments pushing the limits on a range of precision observables, such as the Hubble constant, the dark energy equation of state, the tensor-to-scalar ratio imprinted on the cosmic microwave background, and the effective number of relativistic neutrino degrees of freedom. On the theory side, these processes are all governed by Boltzmann equations, which describe the evolution of interacting particle distributions in an expanding universe. Existing Boltzmann solvers make the underlying equations tractable through combinations of momentum discretisation, assumptions about the form of phase-space distributions, and problem-specific reductions of microscopic interaction rates. Parameter scans with conventional solvers are also expensive, often prohibitively so, as the full system must be re-solved for each parameter combination. The upcoming precision targets, together with the growing interest in new physics scenarios, call for Boltzmann solvers that overcome these computational limitations.

We introduce Neural Boltzmann Equations (NBEs), a framework for solving early-universe Boltzmann equations. Our approach focuses on flexibility without extensive process-specific assumptions, and scalability to complex and compute-intensive problems. NBEs combine three concepts:
\begin{itemize}
    \item Neural distribution functions -- a physics-inspired parameterisation that includes equilibrium solutions, works without discretisation, and can be sampled efficiently. The parameters can be predicted with a neural network based on arbitrary conditioning variables, enabling parameter scans with a single distribution function solve.
    \item Monte Carlo collision integrals -- high-dimensional phase-space integrals, computed with importance sampling techniques developed for event generators at colliders.
    \item Natural gradient training -- an efficient method of training neural distribution functions on the Boltzmann equation to high precision.
\end{itemize}
We discuss the three aspects in detail in Section~\ref{sec:method}. We then use toy problems to demonstrate the individual benefits of NBEs in Section~\ref{sec:results}. Finally, we use the framework to perform a precision calculation of the effective number of relativistic neutrino degrees of freedom in the Standard Model, and show that an NBE trained once can perform a full parameter scan for a new physics model.

\section{Background} \label{sec:background}

\textbf{Early-universe dynamics.} 
The Boltzmann equation governs the evolution of particle populations in spacetime, coupling microscopic interactions to propagation on a non-trivial background geometry,
\begin{equation}\label{eq:BoltzmannEq}
    \mathcal{L}[f_a] = \mathcal{C}_a[f]\;,
\end{equation}
where $f_a\ge 0$ is the dimensionless distribution function of species $a$, representing the occupation of its phase space. The exchange symmetry of identical particles distinguishes two particle types: bosons where $f_a>1$ is possible, and fermions which satisfy $f_a\le 1$. The Liouville operator of species $a$, $\mathcal{L}[f_a]$, describes the propagation of particles in curved spacetime, whilst the collision operator, $\mathcal{C}_a[f]$, encodes particle interactions involving all contributing species $f=(f_1,f_2,\dots)$. 

We specialise to FLRW spacetime~\citep{Friedman1922,1927ASSB...47...49L,1935ApJ....82..284R,WalkerFLRW}, which captures the observed expansion, large-scale homogeneity, and isotropy of the universe. Homogeneity and isotropy reduce the dependence of the distribution functions $f_a$ on 3 spatial and 3 momentum variables to a single momentum variable. Expansion is characterised by the scale factor, $a(t)$, and naturally absorbed by working in comoving variables. In terms of the comoving evolution variable $x = a m_0$, with $m_0$ the characteristic energy scale, the Liouville operator is
\begin{equation}
    \mathcal{L}[f_a] = Hx \frac{\partial f_a}{\partial x}\;, \qquad H = \frac{x}{m_0} \frac{\dot a}{a}\;,
\end{equation}
with $H$ the comoving Hubble parameter~\citep{Hubble:1929ig}. More generally, any physical quantity $Q^{\mathrm{phys}}$ with mass dimension $d$ has a comoving counterpart $Q = (x/m_0)^d Q^{\mathrm{phys}}$, and we fully work in comoving quantities from now on. We reserve the symbol $y$ for comoving momenta. This places all relevant evolution near $x=1$, irrespective of the physical problem.

The collision operator details all processes that populate or deplete the phase space of species $a$. For each process $k$, the statistical factor $\Lambda_k[f]$ combines incoming and outgoing distribution functions with Pauli blocking, ($1-f_b$), and Bose enhancement, ($1+f_b$), factors for outgoing fermions or bosons. The process contribution is weighted by a squared matrix element $|\mathcal{M}_k|^2$ quantifying the interaction probability, and integrated over the phase space of all particles in the process $k$ except particle $a$, denoted as $k\setminus a$ in $\dd\Phi_{k\setminus a}$, using the phase space measure $\dd\Phi_k$ of~\eqref{eq:phasespace},
\begin{align}\label{eq:collision-operator}
\mathcal{C}_a[f] &= \frac{1}{2E_a}\sum_k \int \dd\Phi_{k\setminus a}\, |\mathcal{M}_k|^2 \Lambda_k[f]\;,\quad \Lambda_k [f] = \prod_{b\in k_\mathrm{out}}f_b \prod_{b\in k_\mathrm{in}}(1\mp f_b) - \left(\mathrm{in}\leftrightarrow\mathrm{out}\right)\;.
\end{align}
Here $E_a=(m_a^2+y^2)^{1/2}$ and $m_a$ are the particle energy and mass, respectively. Thermal equilibrium is characterised by detailed balance, $\Lambda_k[f] =0$ for each process and phase-space point, implying $\mathcal{C}_a[f]=\mathcal{L}[f_a] =0$. Solving $\Lambda_k[f]=0$ yields the equilibrium solutions $f_\mathrm{eq}(E) = [\exp((E-\mu)/T)\pm 1]^{-1}$, the Fermi-Dirac ($+$) and Bose-Einstein ($-$) distributions, parameterised by the temperature $T$ and chemical potential $\mu$.
Macroscopic observables such as number density $n_a$, energy density $\rho_a$, and pressure $P_a$ are moments of the distribution function: $M_a[\phi] = \int \dd^3y/(2\pi)^3 \phi (y)f_a(y)$. Specifically, $n_a=M_a[1]$, $\rho_a=M_a[E]$, and $P_a=M_a[y^2/(3E)]$.


\textbf{Phase-space integrals.}
Predictions in high-energy physics, such as scattering rates at colliders, decay widths of unstable particles, or reaction rates in the early universe, require integration over the kinematics of all particles. The central object is the Lorentz-invariant phase-space measure
\begin{equation}\label{eq:phasespace}
    \dd\Phi_k = \left(\prod_{a\in k} \dd \Pi_a\right) (2\pi)^4 \delta^4 \left(\sum_{a\in k_\mathrm{in}}y_a-\sum_{a\in k_\mathrm{out}}y_a\right) \;,
    \qquad \dd\Pi_a = \frac{\dd^3 y_a}{(2\pi)^3 2E_a}\;.
\end{equation}
The one-particle Lorentz-invariant phase-space measure, $\dd\Pi_a$, is included for every particle that is integrated over, and the delta distribution, $\delta^4 (\,\cdot\,)$ enforces conservation of both energy, $E_a$, and spatial momentum, $\vec y_a$, via the four-momentum $y_a=(E_a,\vec y_a)$. This leaves a $(3n-4)$-dimensional integral for $n$ integrated particles. For $n>2$, Monte Carlo integration is the only efficient  integration method. After mapping integration variables to the unit hypercube $u\in [0,1]^{3n-4}$, the integrals read
\begin{equation}\label{eq:mc-integration}
    I = \int \dd u\,h(u) = \mathbb{E}_{u\sim p(u)}\left[\frac{h(u)}{p(u)}\right]\;,\qquad \sigma_I^2 = \frac{1}{N}\mathrm{Var}_{u\sim p(u)}\left[\frac{h(u)}{p(u)}\right]\;.
\end{equation}
The proposal distribution, $p(u)$, should be chosen to approximate the normalised integrand. The variance, $\sigma_I^2$, measures the approximation quality, and vanishes for $p(u) = h(u)/I$, $h(u)\ge 0$. 

High-energy physics event generators use several techniques to improve the quality of the proposal distribution, $p(u)$~\citep{Alwall:2011uj,Sherpa:2019gpd}. Process-specific phase-space mappings map the unit hypercube to the space of valid momentum configurations, accounting for momentum conservation and the integrand shape.
More recently, normalising flows and other generative networks have been used to further refine the proposal distribution~\citep{Heimel:2022wyj}.

\textbf{Natural gradient optimisation.}
The natural gradient method makes use of the function space geometry to enable optimal parameter updates~\citep{Amari:2000ig}. This method is becoming increasingly popular in Variational Monte Carlo~\citep{Stokes:2019pmg} and physics-informed neural networks~\citep{Mueller:2023engd}. Consider the problem of fitting the function $g(y;\theta)$, with $\theta\in \mathbb{R}^N$, to the target function $g_\mathrm{target}(y)$, with residuals weighted by the factor $W=1/g(\theta)$,
\begin{equation}\label{eq:natgrad-setup}
    \theta_* = \mathrm{argmin}_\theta L(\theta)\;,\qquad L(\theta) = r^\top Wr\;,\qquad r=g(\theta) - g_\mathrm{target}\;.
\end{equation}
We sample $B$ values of $y$ from a density $p(y)$, where we evaluate $g$, $g_\mathrm{target}$ and $p$, and build the vectors $g(\theta),g_\mathrm{target},p\in\mathbb{R}^B$. The density factor $p$ enters the weighting matrix $W=\mathrm{diag}(1/(p\, g(\theta)))$. To solve the problem, we linearise the system as $g(\theta )=g(\theta_0) +J(\theta-\theta_0)$ with Jacobian $J=\partial g/\partial\theta$. The optimal parameter update is then given by the inverse of the Fisher matrix $F\in\mathbb{R}^{N\times N}$
\begin{equation}\label{eq:natgrad-step}
    \theta_* = \theta_0 - F^{-1}J^\top W r_0\;,\qquad F=J^\top WJ\;,\qquad r_0=g(\theta_0)-g_\mathrm{target}\;.
\end{equation}
Standard gradient descent neglects the function space geometry and sets $F\propto 1$. 
For small networks, $N\le B$, we invert $F$ directly, whereas for large networks, $N> B$, we follow \citet{Chen:2023njx} and rewrite $(J^\top WJ)^{-1}J^\top W = J^\top (JJ^\top)^{-1}$, then invert the Gram matrix $JJ^\top\in\mathbb{R}^{B\times B}$ instead.

\section{Related work}

\textbf{Boltzmann equation solvers.} 
Existing solvers are typically tailored to a specific problem, each with its own analytic reductions and numerical approximations to make its system tractable. For neutrino decoupling, FortEPiaNO evolves neutrino density matrices on a fixed Gauss-Laguerre momentum grid with analytically reduced collision integrals~\citep{Gariazzo:2019gyi}. Neutrino direct simulation Monte Carlo~\citep{Ovchynnikov:2024rfu,Ovchynnikov:2024xyd} instead tracks individual neutrinos and simulates interactions with Monte Carlo. 
For the cosmic microwave background (CMB), CAMB and CLASS solve the Einstein-Boltzmann hierarchy on a fixed background,  predicting CMB anisotropies and the matter power spectrum~\citep{Lewis:1999bs,Blas:2011rf}. 
For dark matter, micrOMEGAs computes relic abundances for generic models using momentum-integrated Boltzmann equations, where thermally averaged rates replace collision operators~\citep{Belanger:2018ccd,Alguero:2023zol}. DRAKE goes beyond kinetic equilibrium and solves for the distribution function of non-relativistic, dilute dark matter in an equilibrium background~\citep{Binder:2021bmg}. ULYSSES specialises to leptogenesis and primarily solves for moments of the distribution function in an equilibrium Standard Model bath with analytically reduced collision integrals~\citep{Granelli:2020pim,Granelli:2026goh}.

NBEs address two bottlenecks of existing Boltzmann solvers. First, moment-based solvers discard phase-space information, while momentum-grid solvers scale steeply with grid size, \textit{e.g.} cubically for $2\leftrightarrow 2$ processes. Parameter scans multiply this cost, since each mass or coupling requires a full solve. NBEs instead parameterise distribution functions with the physics-inspired neural distribution function ansatz, and conditioning on masses or couplings allows one model to learn a family of solutions across parameter space. Second, beyond $2\leftrightarrow 2$ processes, collision integrals become prohibitively expensive with standard quadrature methods: every additional particle adds a three-dimensional momentum integral, so that analytic reductions and grid choices must be re-derived for each new process. NBEs instead use Monte Carlo methods, which scale well to complex processes and require only the matrix element for a new process, with no fixed grids or analytic reductions.

\textbf{Neural quantum states.}
A neural network serves as a variational ansatz for the state of a quantum system, trained directly on the equations governing the system rather than data. In the original formulation for quantum lattice models, a network parameterises the many-body wave function $\psi (s;\theta)$ for discrete lattice configurations $s$, and the parameters $\theta$ are optimised using the variational principle~\citep{Carleo:2016svm}. Neural wave functions~\citep{Pfau:2019,Hermann:2019} extend this idea to electrons in continuous space, solving the electronic Schrödinger equation of atoms and molecules from first principles. Neural distribution functions do the same for kinetic theory: a network parameterises phase-space distribution functions and is trained on the residual of the Boltzmann equation, which plays the role of the Hamiltonian in the examples above.

\section{Methods}
\label{sec:method}

Neural Boltzmann Equations combine three ingredients. First, we construct a physics-inspired ansatz for bosonic and fermionic distribution functions that generalises the known equilibrium solutions. Second, we efficiently evaluate phase-space integrals using Monte Carlo integration with importance sampling. Third, we solve the Boltzmann equation sequentially, using natural gradient updates at each time step. Additional details are given in Appendix~\ref{app:method}.

\subsection{Neural distribution functions}
\label{sec:neural_distribution_functions}

Thermodynamic distribution functions detail the mean occupation at each phase-space point. Motivated by this property, we factorise distribution functions into the invariant density $\eta$, which sets the overall scale, and the normalised density, $p(y)$, encoding their momentum distribution:
\begin{equation} \label{eq:distribution-function}
    f (y) = \frac{4\pi^2E}{y^2} \eta p (y)\;,\qquad \eta = \int \dd y\,\frac{y^2}{4\pi^2E}f (y)\;,\qquad \int \dd y\, p(y)=1\;.
\end{equation}
The parameterisation uses the invariant density $\eta$ instead of the number density, $n$, because it naturally appears in phase-space integrals, making $p(y)$ the optimal importance sampling distribution of the Lorentz-invariant phase space, see Appendix~\ref{app:collision}. The deterministic prefactor $4\pi^2 E/y^2$ ensures that the second equation above holds. 
We parameterise the normalised density with a mixture of gamma distributions, $\gamma (\varepsilon | \alpha, \lambda)$, in terms of the rescaled kinetic energy $\varepsilon = (\sqrt{y^2+m^2}-m)/T$ as
\begin{equation} \label{eq:gamma-mixture}
    p(y) = \frac{y}{ET} \sum_{m=1}^M w_m\gamma \left(\frac{\sqrt{y^2+m^2}-m}{T} \Big| \alpha_m, \lambda_m\right)\;,\qquad \gamma (\varepsilon  |\alpha, \lambda )=  \frac{\lambda}{\Gamma (\alpha )}(\lambda \varepsilon)^{\alpha -1} e^{-\lambda \varepsilon}\;.
\end{equation}
With sufficiently many mixture components $M$, this parameterisation can represent the massless bosonic equilibrium distribution to arbitrary precision, see Appendix~\ref{app:distribution-function}. Fermionic distribution functions, which must satisfy the Pauli exclusion principle, $f(y)\le 1$, are constructed from an unconstrained bosonic $f_0(y)$,
\begin{equation}\label{eq:fermions-distribution-function}
    f(y) = \frac{f_0(y)}{1+f_0(y)}\in [0, 1]\;,\qquad p(y) = \frac{1}{N}\frac{1}{1+f_0(y)}p_0(y)\;,\qquad N=\int \dd y\, \frac{1}{1+f_0(y)}p_0(y)\;.
\end{equation}
The Fermi-Dirac distribution corresponds to $f_0(y)=e^{-(E-\mu)/T}$, a single mixture component. We can sample from the normalised densities, $p(y)$, for both bosons and fermions, although the fermionic case requires an additional rejection sampling step with typical acceptance rate $0.8$, see Appendix~\ref{app:distribution-function}. Moments of $f(y)$, such as number and energy density, can be evaluated efficiently using Gauss-Laguerre or Gauss-Hermite quadrature. See Appendix~\ref{app:distribution-function} for additional information.

Together, the invariant density, $\eta$, and gamma mixture parameters, $(w_m,\alpha_m,\lambda_m)_{m=1}^M$, fully specify the parameterisation $\omega$, which has $3M$ independent parameters due to the redundancy in the normalisation of the mixture weights. Depending on the application, $\omega$ is either optimised directly or predicted from theory conditions, $\xi$, by a neural network with weights $\theta$,
\begin{equation}
    \omega = \left\{\eta, \left(w_m, \alpha_m, \lambda_m\right)_{m=1}^M\right\} = \mathrm{NN}(\xi;\theta)\;.
\end{equation}
We write $f(y;\omega)$ and $p(y|\omega)$ to highlight the dependence on the parameters $\omega$, and use $M=50$ throughout. If predicted with a neural network, the parameterisation is cached and reused after the initial evaluation. The same parameterisation describes both unconstrained and equilibrium distribution functions, where only temperature and chemical potential are kept as free parameters.

\subsection{Collision operator}
\label{sec:collision_integrals}

The collision operator, $\mathcal{C}_a[f]$, quantifies the net change in particle occupations due to microscopic interactions. We give the explicit expression for $\mathcal{C}_a[f]$ in \eqref{eq:collision-operator}, and the phase-space measure $\dd \Phi_k$ in \eqref{eq:phasespace}. Evaluating this integral is typically the compute bottleneck in the NBE framework.

\begin{figure}[t]
\begin{algorithm}[H]
  \caption{Phase-space generation for a process with $m$ incoming and $n-m$ outgoing particles.}
  \label{alg:collision-sampling}
  \begin{algorithmic}[1]
    \Require observed momentum $y_1$, condition $\xi$, masses $m_1,\dots,m_n$, proposal densities $p_2,\dots,p_m$
    \State $y_k \sim p_k(\,\cdot \mid \xi)$, $p_k\gets p_k(y_k|\xi)$ for $k=2,\dots,m$ \Comment{sample initial momenta}
    \State $\;\hat\Omega_k \sim \mathcal U(S^2)$ for $k=1,\dots, m$ \Comment{sample initial angles}
    \State $u_{\rm out} \sim \mathcal U\big([0,1]^{3(n-m)-4}\big)$ \Comment{sample outgoing configuration}
    \If{neural importance sampling}
      \State $u,\, J_{\rm enc} \gets \operatorname{ToUnitCube}(y_2,\dots,y_m,\hat\Omega_2,\dots,\hat\Omega_m, u_\mathrm{out})$
      \State $u,\, J_{\rm flow} \gets \operatorname{Flow}(u \mid y_1, \xi)$
      \Comment{MadNIS refinement}
      \State $(y_2,\dots,y_m,\hat\Omega_2,\dots,\hat\Omega_m,\, u_{\rm out}),\, J_{\rm dec} \gets \operatorname{ToUnitCube}^{-1}(u)$
    \Else
      \State $J_{\rm enc},\, J_{\rm flow},\, J_{\rm dec} \gets 1$
    \EndIf
    \State $\vec y_k \gets y_k \hat\Omega_k$ for $k=1,\dots,m$ \Comment{build initial state}
    \State $\sqrt s \gets \big[(\textstyle\sum_{k\le m} E_k)^2 - |\sum_{k\le m} \vec y_k|^2\big]^{1/2}$
           with $E_k = (y_k^2 + m_k^2)^{1/2}$
    \State $\vec y_{m+1},\dots,\vec y_n,\, J_{\rm ps} \gets \operatorname{MadSpace}(u_{\rm out};\, \sqrt s,\, m_{m+1},\dots,m_n)$
           \Comment{build final state}
    \State $w_\mathrm{ps} \gets J_{\rm enc}\, J_{\rm flow}\, J_{\rm dec}\, J_{\rm ps} \prod_{k=2}^{m}
           y_k^2 / (4\pi^2 E_k p_k)$ \Comment{importance weight}
    \State \Return $(\vec y_1,\dots,\vec y_n;\, w_\mathrm{ps})$
  \end{algorithmic}
\end{algorithm}
\vspace{-0.8cm}
\end{figure}

To evaluate the collision operator with Monte Carlo integration, we first generate phase-space points following Algorithm~\ref{alg:collision-sampling}. The initial-state angles, $\hat\Omega_k$, are sampled uniformly on the sphere and the momenta, $y_k$, from the neural distribution function densities, $p_k$. For the final state, we sample $u_\mathrm{out}$ uniformly from the unit hypercube and use MadSpace~\citep{Heimel:2026hgp} to map it onto valid phase-space configurations, with the initial-state information entering through the centre-of-mass energy, $\sqrt{s}$.
We use neural importance sampling to further refine the phase-space generation with a normalising flow. The deterministic $\mathrm{ToUnitCube}$ transform maps all generated variables to the unit hypercube, where we apply the learned flow transform. We use the MadNIS implementation for the normalising flow~\citep{Heimel:2023ngj,Heimel:2026hgp}, conditioned on the fixed momentum, $y_1$, parameters, $\xi$, and trained to minimise the variance of the integral, see \eqref{eq:madnis-loss}. The framework also accommodates advanced event-generation techniques such as topology-aware phase-space mappings and multi-channel Monte Carlo, though we do not need them for the examples in this work.

Phase-space generation returns the importance weight, $w_\mathrm{ps}$, and the momentum configurations, $\vec y_1,\dots, \vec y_n$, which determine the matrix element, $|\mathcal{M}_k|^2$, and the gain-loss combination, $\Lambda_k[f]$. Summing over processes yields the collision operator
\begin{equation}
    \mathcal{C}_a[f](y_1) = \frac{1}{2E_a} \sum_k \int \dd\Phi_{k \setminus a}|\mathcal{M}_k|^2 \Lambda_k [f] = \frac{1}{2E_a}\sum_k \mathbb{E}_{\hat\Omega_1,\vec y_2,\dots , \vec y_n}\left[w_\mathrm{ps} |\mathcal{M}_k|^2 \Lambda_k [f]\right]\;.
\end{equation}
Precise evaluations of the collision integral demand large sample counts, and so dominate the NBE compute cost. We mitigate this by parallelising across particles and processes, and using the optimised MadSpace library. As the training procedure in Section~\ref{sec:training} passes no gradients through the collision integral, we are free to include other event-generation tools if required.

\subsection{Boltzmann equation solver}
\label{sec:training}

\begin{figure}[t]
\begin{algorithm}[H]
  \caption{Boltzmann solver step $\log x_n\to\log x_{n+1}$ fixed-point iteration}
  \label{alg:time-step}
  \begin{algorithmic}[1]
    \Require condition prior $p_\mathrm{prior}(\xi)$, frozen previous solutions $\theta_n$, $\theta_{n-1}, \dots$
    \State $\theta_{n+1}\gets \theta_n$ \Comment{warm-start from previous step}
    \Repeat
      \State $\xi_i \sim p_\mathrm{prior}(\xi)$ for $i=1,\dots,B$ \Comment{sample conditions}
      \State $\xi\gets (\xi_1,\dots ,\xi_B)$ \Comment{batched evaluation}
      \State $\omega_l\gets \mathrm{NN}(\xi; \theta_l)$ for $l=n+1,n,\dots$ \Comment{cache distribution function parameters}
      \For{species $a = 1,\dots,N$} \Comment{update each species independently}
        \State $y \sim p(\,\cdot\, | \omega_{n+1,a})$ \Comment{sample test momenta}
        \State $s(y),\, \kappa(y)\gets $ evaluate collision integral following Section~\ref{sec:collision_integrals}
        \State $A(y),\, \gamma \gets$ evaluate scheme coefficients using Appendix~\ref{app:training}
        \State $g_\mathrm{target}(y) \gets T(g(y;\omega_{n+1,a}); s(y), \kappa (y), A (y), \gamma)$ \Comment{fixed-point target \eqref{eq:fixed-point}}
        \State $r\gets g(y;\omega_{n+1,a}) - g_\mathrm{target}(y)$
        \State $w \gets \big[ p(y |\omega_{n+1,a})\, g(y;\omega_{n+1,a}) \big]^{-1}$
        \State $J\gets \frac{\partial g(y;\omega (\theta))}{\partial\theta}\Big\rvert_{\theta=\theta_{n+1,a}}$ using autodiff
        \State $W\gets \mathrm{diag} (w)$, $F\gets J^\top WJ$
        \State $\theta_{n+1,a} \gets \theta_{n+1,a} - F^{-1} J^\top W r$ \Comment{natural-gradient step \eqref{eq:natgrad-step}}
      \EndFor
    \Until{objective $r^\top Wr$ plateaus or Monte Carlo noise dominates}
  \end{algorithmic}
\end{algorithm}
\vspace{-0.8cm}
\end{figure}

We now discuss the training of neural distribution functions to satisfy the Boltzmann equation. Multiplying \eqref{eq:BoltzmannEq} by $y^2/H$, writing $g(y;\theta )=y^2f(y;\theta)$ for the fit function, $s=y^2 \mathcal{C}/H$ for the source term, and dropping the particle index $a$, the Boltzmann equation reads
\begin{equation}\label{eq:simple-boltzmann}
    s(g) = \frac{\partial g}{\partial\log x} \simeq \frac{g_{n+1}-A}{\gamma h}\;,
\end{equation}
where in the rightmost expression we discretised the derivative with step size $h=\Delta\log x$, and defined $g_m(y)\equiv g(y;\theta_m)$ for the fit function at step $m$. The discretisation scheme $(A,\gamma)$ combines $A=A(g_n, g_{n-1},\dots)$, which depends on the previous solutions, with a step-size weight $\gamma$. The standard difference quotient is $(A,\gamma)=(g_n,1)$. We use a multi-step scheme, see Appendix~\ref{app:training}.

To solve \eqref{eq:simple-boltzmann}, we consider the implicit equation $0=F(g_{n+1}) = g_{n+1}-A-\gamma h s(g_{n+1})$, solved by iterating $g_{n+1}' = T(g_{n+1}) = g_{n+1} - M^{-1}F(g_{n+1})$ with an arbitrary function $M$. The choice $M=\partial F/\partial g=1-\gamma h\,\partial s/\partial g$ converges fastest, but requires a costly collision operator derivative. We instead use $M=1 + \gamma h \kappa$, with $\kappa\ge 0$ a cheap approximation of $-\partial s/\partial g$ that exploits the specific form of the collision operator, see Appendix~\ref{app:training}. In contrast to the naive choice, $M=1$, this yields an update rule that is invariant under small perturbations of $g_{n+1}$, making the formalism applicable to stiff regimes that commonly arise whenever species are deep in thermal equilibrium,
\begin{equation}\label{eq:fixed-point}
    g_{n+1}' = T(g_{n+1}) = \frac{A+\gamma h (s(g_{n+1}) + \kappa g_{n+1})}{1+\gamma h\kappa}\;.
\end{equation}

Algorithm~\ref{alg:time-step} summarises the training for a single time step. Using the previous step parameters as an initial guess, $\theta_{n+1}=\theta_n$, we iterate until the objective plateaus or the feedback is dominated by Monte Carlo noise, see Appendix~\ref{app:training}. Each iteration samples a batch of conditions, $\xi$, caches the neural distribution function parameters, $\omega_{n+1}$, $\omega_n$, etc., for the required time steps, and updates the parameters $\theta_{n+1,a}$ of each species $a$ independently. After sampling test momenta from the distribution function density, we evaluate the collision integral objects: $s$, $\kappa$, the scheme coefficients, $A$, and $\gamma$. The iteration target follows from \eqref{eq:fixed-point}, and a natural gradient step moves the parameters $\theta_{n+1,a}$ towards it. We use $\mathcal{O}(10^3)$ test points per particle, and $\mathcal{O}(10^3)$ collision integral evaluations per test point. The procedure converges within 5-10 (10-30) iterations for unconditional (conditional) distribution functions, depending on the sensitivity to the condition. To enforce the initial condition $g(y;\theta_0) = g_\mathrm{init}(y)$, we use the same iterative algorithm with deterministic target $g_\mathrm{target}(y)=g_\mathrm{init}(y)$, which typically converges within 50 iterations. See Appendix~\ref{app:training} for details.

For neural importance sampling, we pretrain the normalising flow on the initial distribution function and evolve it jointly with the neural distribution functions. In each iteration, we use one collision integral sample per $(y_1,\xi)$ pair to update the normalising flow density $q(u|y_1,\xi)$, with $q(u|y_1,\xi)\rvert_\text{no-grad}$ denoting terms excluded from the gradient calculation. We train the flow on a variance loss using the Adam optimiser, since natural gradient training gave no significant benefit here:
\begin{equation}\label{eq:madnis-loss}
    \mathcal{L} = \mathbb{E}_{\xi\sim p_\mathrm{prior}(\xi),y_1\sim p(y_1|\xi),u\sim q(u|y_1,\xi)}\left[\left(\frac{s(u)}{p(y_1|\xi) q(u|y_1,\xi)}\right)^2\Big\rvert_\text{no-grad} \frac{q(u|y_1,\xi)\rvert_\text{no-grad}}{q(u|y_1,\xi)}\right]\;.
\end{equation}

\textbf{Computational cost.} 
The collision operator dominates the cost, with typically $10^3$ function evaluations per test momentum, whereas the network evaluation, natural gradient update, and MadNIS update do not scale with the collision operator sample count, and only contribute at the $10\%$ level. The training runs in Section~\ref{sec:results} take up to 5 hours on an NVIDIA A30 GPU, see Appendix~\ref{app:experiment-details}.

\textbf{Limitations.} 
NBEs use a parametric ansatz for the distribution function, which may require many components to express exotic shapes. Adding different components to the gamma mixture model may improve efficiency, but requires knowledge of the specific effects. NBEs also assume homogeneous, isotropic FLRW spacetime, reducing distribution function dependence from $f(\vec y,\vec x)$ to $f(y)$.

\section{Results} \label{sec:results}

Here we showcase the properties of NBEs on toy problems, and then perform a precision calculation of the effective number of relativistic neutrino degrees of freedom in the early universe. See Appendix~\ref{app:experiment-details} for details on the experiments, and Appendix~\ref{app:additional-results} for additional results.

\subsection{Equilibrium distributions} \label{sec:detailed balance}

\begin{figure}[t]
    \centering
    \includegraphics[width=0.495\textwidth,page=1]{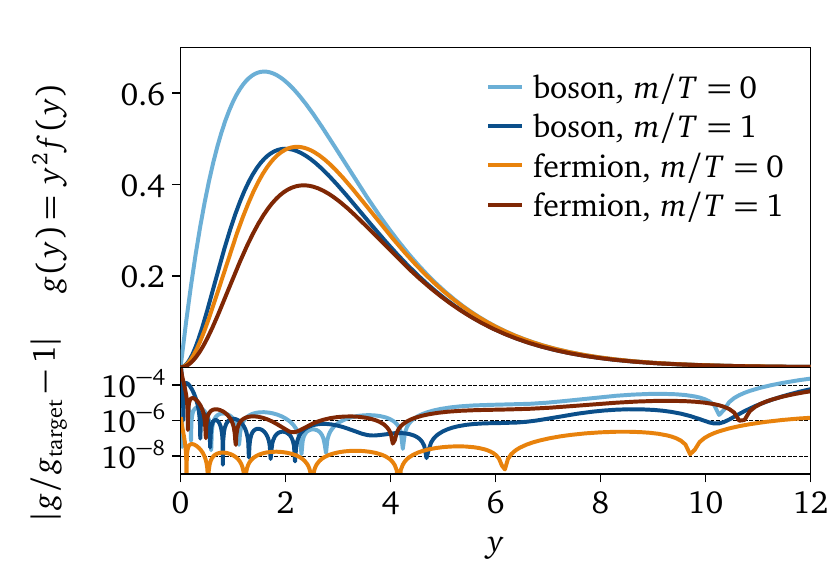}
    \includegraphics[width=0.495\textwidth,page=2]{figs/equilibrium.pdf}
    \vspace{-0.8cm}
    \caption{\textbf{Natural gradient training for neural distribution functions.} Left: Learned distribution functions for massless and massive bosons and fermions, compared to equilibrium targets. Right: Average residual over training time for massless bosons and fermions, comparing the natural gradient method with the Adam optimiser using the mean and standard deviation of ten independent trainings.}
    \label{fig:equilibrium}
\end{figure}

As a first test of the NBE framework, we train neural distribution functions with Algorithm~\ref{alg:time-step} on a fixed target, $g_\mathrm{target}(y)=g_\mathrm{init}(y)$, i.e. the initialisation before time evolution. The left panel of Figure~\ref{fig:equilibrium} shows the learned distribution functions for massive and massless bosons and fermions, which match equilibrium to $10^{-6}$ relative deviation in the bulk. The massless fermion distribution is exactly representable by a single mixture component in our parameterisation and therefore agrees to even higher precision. The precision is limited by the restriction to 50 mixture components, and by regularisation effects in the Fisher matrix inversion.

Second, we compare natural gradient training with standard gradient descent using Adam~\citep{Kingma:2014vow} in the right panel of Figure~\ref{fig:equilibrium}. Natural gradient training converges to an average residual of $10^{-6}$ within 10 iterations, far outperforming Adam at $10^{-2}$ for the same iteration budget. The Adam residual continues to fall and reaches $10^{-4}$ after around 10,000 iterations. Although a natural gradient update takes about twice as long as an Adam update in our implementation, it is far more time-efficient overall and reaches a higher final precision thanks to the curvature information.

\subsection{Thermalisation} \label{sec:thermalisation}

\begin{figure}[t]
    \centering
    \includegraphics[width=0.495\textwidth,page=1]{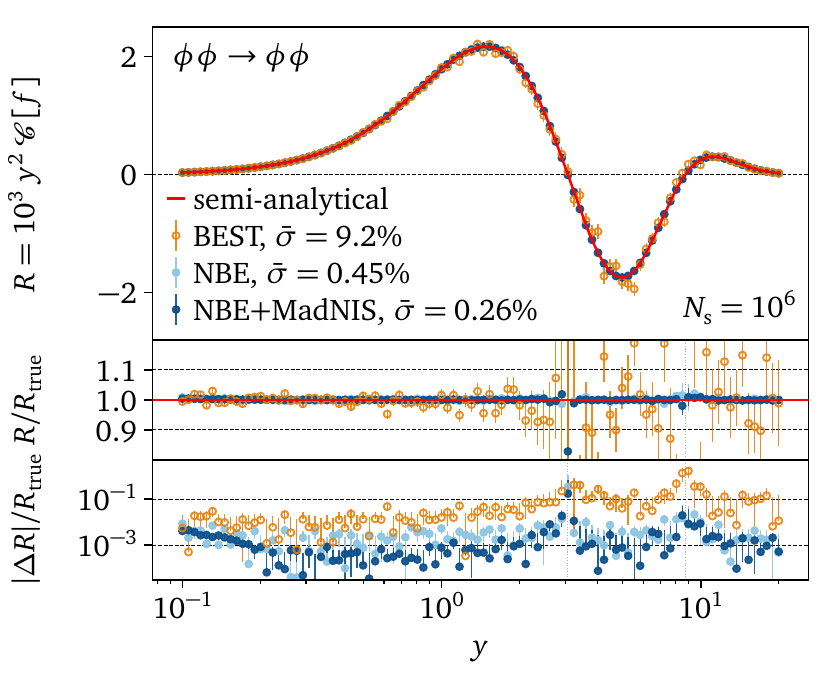}
    \includegraphics[width=0.495\textwidth,page=1]{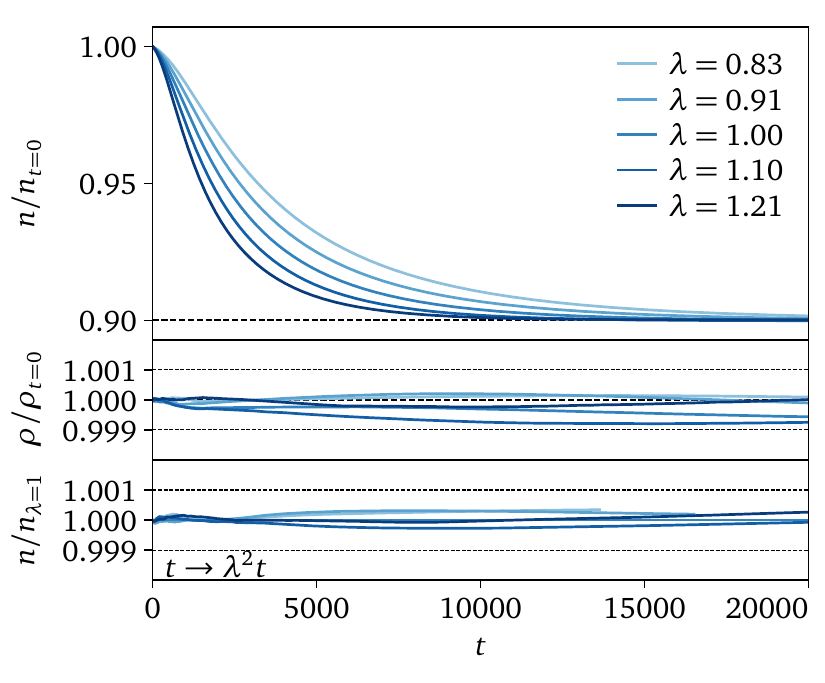}
    \vspace{-0.8cm}
    \caption{\textbf{Efficient collision operator evaluation and coupling-conditioned training.} Left: Collision operator for $\phi\phi\leftrightarrow\phi\phi$, calculated semi-analytically and with Monte Carlo using BEST and NBE, with the momentum-averaged relative Monte Carlo uncertainty, $\bar\sigma$, as precision metric. Right: Number density, $n_t$, and energy density, $\rho_t$, over time for different couplings, see \citet{Yoon:2026rce} for the BEST result. The lower panel rescales $t\to\lambda^2 t$ to collapse the curves for different couplings.}
    \label{fig:thermalisation}
\end{figure}

We now consider a single massive scalar particle without cosmic expansion~\citep{Yoon:2026rce}, a toy, strongly interacting massive particle dark matter model~\citep{Hochberg:2014dra}. Starting out of equilibrium, the particle thermalises via the processes $\phi\phi\leftrightarrow\phi\phi$ and $\phi\phi\leftrightarrow\phi\phi\phi$ with common coupling strength $\lambda$. Whereas the $\phi\phi\leftrightarrow\phi\phi$ collision integral is 2-dimensional, the $\phi\phi\leftrightarrow\phi\phi\phi$ integral is 5-dimensional, motivating Monte Carlo integration.

The left panel of Figure~\ref{fig:thermalisation} shows the collision operator as a function of the momenta, $y$, using NBEs, and the Boltzmann Equation Solver for Thermalisation (BEST)~\citep{Yoon:2026rce}, which uses a custom phase-space mapping refined with the VEGAS algorithm~\citep{Lepage:2020tgj}. The momentum-averaged Monte Carlo uncertainty, $\bar\sigma$, is $20\times$ smaller for NBE than for BEST, corresponding to a $400\times$ more sample-efficient estimate, given that $\bar\sigma\propto 1/\sqrt{N_s}$. This gain comes from the optimised importance sampling, namely MadSpace phase-space mappings and distribution function sampling. MadNIS refinement increases the gain to $1000\times$ in total.

We next condition an NBE on the coupling $\xi = \lambda$ with a uniform prior and evolve the system in time. The energy density $\rho_t$ is conserved, whilst number-changing processes drive the number density $n_t$ to $n_{t=\infty}/n_{t=0}\approx 0.900$ in equilibrium. The right panel of Figure~\ref{fig:thermalisation} demonstrates the number density converging to the asymptotic value at a coupling-dependent rate, whilst the energy density remains constant to within $0.1\%$. Rescaling $t\to \lambda^2t$ should align all curves, see Appendix~\ref{app:experiment-details}, and NBEs realise this to $< 0.1\%$. Although it must cover the full prior range, this conditional solve is only about $5\times$ slower than a comparable fixed-coupling solve in our implementation.

\subsection{Neutrino decoupling} \label{sec:Neff}

\begin{figure}[tb]
    \centering
    \includegraphics[width=0.24\textwidth,page=1]{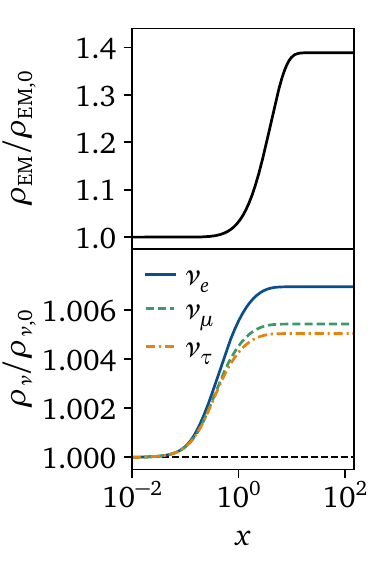}
    \includegraphics[width=0.24\textwidth,page=2]{figs/neff_evolution.pdf}
    \includegraphics[width=0.49\textwidth,page=1]{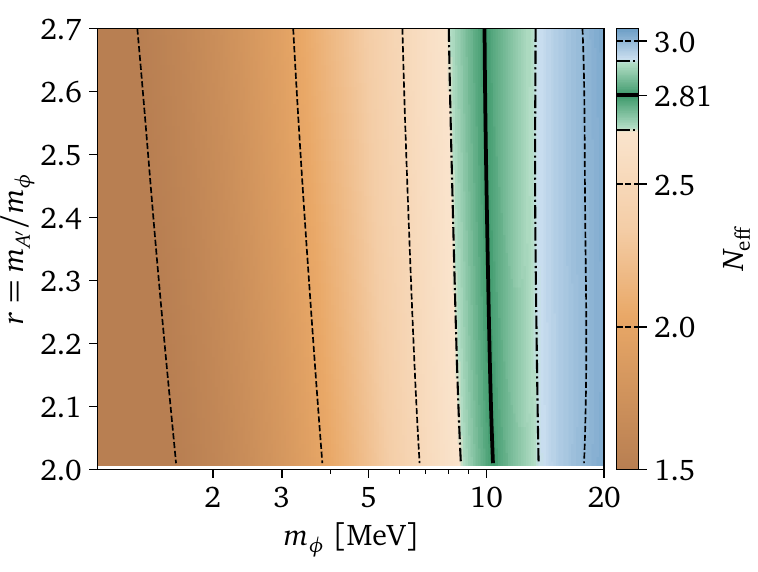}
    \vspace{-0.5cm}
    \caption{\textbf{Neutrino decoupling and new physics impact.} Left: Evolution of electromagnetic and neutrino energy densities during neutrino decoupling. Middle: Neutrino spectral distortions after decoupling, comparing the equilibrium solution to the full solution including spectral distortions. Right: $N_\mathrm{eff}$ for the new physics model of \citet{Escudero:2026mgw}, from a single NBE training.}
    \label{fig:neff}
\end{figure}

In the early universe, weak interactions keep neutrinos in equilibrium with the electromagnetic plasma of photons ($\gamma$), electrons ($e^-$), and positrons ($e^+$) through scattering ($\nu e^\pm\leftrightarrow \nu e^\pm$) and annihilation ($\nu\bar\nu\leftrightarrow e^+e^-$) processes. 
At temperatures of $T\approx 1\,\mathrm{MeV}$, or around one second after the Big Bang, neutrinos lose thermal contact with the electromagnetic plasma due to cosmic expansion, and free stream as the cosmic neutrino background. Shortly afterwards, at $T\approx m_e\approx 0.5\,\mathrm{MeV}$, electrons and positrons annihilate ($e^+e^-\to\gamma\gamma$), reheating photons. The overlap of these processes controls the relative heating of the neutrino and electromagnetic sectors, and therefore the precision observable $N_\mathrm{eff}$, the effective number of relativistic neutrino degrees of freedom:
\begin{equation}
    N_\mathrm{eff} = \frac{8}{7}\left(\frac{11}{4}\right)^{4/3} \frac{\rho_\nu}{\rho_\mathrm{EM}}\Big\rvert_{T/m_e\to 0}\;.
\end{equation}
Entropy conservation yields $N_\mathrm{eff}=3$ by construction for instantaneous decoupling. Interactions increase $N_\mathrm{eff}$ to $3.0440$, see Appendix~\ref{app:neff-theory} for analytic expressions and Appendix~\ref{app:experiment-details} for numerical results on the individual effects. We visualise the process in Figure~\ref{fig:neff}: the electromagnetic energy density $\rho_\mathrm{EM}=\rho_\gamma +\rho_e + \rho_\mathrm{QED}$ increases significantly due to electron-positron annihilation, whereas the neutrino energy densities $\rho_\nu$ increase only slightly due to the overlap of neutrino decoupling and electron-positron annihilation. Electron neutrinos experience more heating due to their additional interactions with the electromagnetic bath. The neutrino distribution functions depart from equilibrium by $\sim 1\%$ due to the same interactions.

The most precise calculation to date finds $N_\mathrm{eff}=3.0440\,\pm 10^{-4}\,(\text{solver})\,\pm 10^{-4}\, (\text{experiment})$ \citep{Bennett:2020zkv,Froustey:2020mcq,Akita:2020szl}, limited by the solver precision and the measured solar neutrino mixing angle (experiment), both contributing at the $10^{-4}$ level. We repeat this calculation using the NBE framework and the recent NuFIT-6.1 updates on experimental inputs~\citep{Esteban:2024eli,NuFIT61}. Our result agrees, with a $50\times$ smaller solver uncertainty:
\begin{equation}\label{eq:neff-result}
      N_\mathrm{eff} = 3.04392
      \;\pm 2\cdot 10^{-6}\, (\text{solver})
      \;\pm 2\cdot 10^{-5}\, (\text{experiment})\;.
\end{equation}
We discuss the calculation in detail in Appendix~\ref{app:experiment-details}. Besides the now leading experimental uncertainty, there is a theory uncertainty from neglected higher-order effects in the neutrino collision integral. Previous partial calculations found corrections ranging from $7\cdot 10^{-4}$~\citep{Cielo:2023bqp} to $2\cdot 10^{-5}$~\citep{Drewes:2024wbw}, turning these effects into the dominating uncertainty. This motivates a full calculation of the higher-order corrections to the neutrino collision integral, to keep the uncertainty on the theory prediction at a level that allows to neglect it in any experimental analysis.

On the experimental side, the most recent cosmic microwave background analysis reports $N_\mathrm{eff}=2.81\pm 0.12$~\citep{AtacamaCosmologyTelescope:2025blo,AtacamaCosmologyTelescope:2025nti,SPT-3G:2025bzu}, with further improvements expected from ongoing experiments~\citep{SimonsObservatory:2018koc}. This $2\sigma$ tension with the theory prediction may indicate either missing systematic uncertainties or new physics. For instance, \citet{Escudero:2026mgw} extend the Standard Model by a dark photon, $A'$, kinetically mixed with the photon and decaying invisibly $A'\to\phi\phi$ to a complex scalar $\phi$ that serves as a dark matter candidate. In a specific mass window, these decays feed the electromagnetic plasma and raise the electromagnetic energy density while evading other cosmological constraints, see \citet{Escudero:2026mgw} for details.

To demonstrate the flexibility of NBEs, we implement the model of \citet{Escudero:2026mgw} as an extension of our calculation, conditioning on the model parameters $\xi = (m_\phi, r)$ and $r=m_{A'}/m_\phi$, with uniform prior. As shown in the right panel of Figure~\ref{fig:neff}, the prediction agrees with the measured $N_\mathrm{eff}=2.81\pm 0.12$ in a band $m_\phi\in [8\,\mathrm{MeV},13\,\mathrm{MeV}]$, nearly independent of the mass ratio $r$.

\section{Conclusion}

We have introduced Neural Boltzmann Equations (NBEs), a flexible framework for solving Boltzmann equations in the early universe. Collision integrals are evaluated with Monte Carlo integration, allowing the framework to scale comfortably to complex processes. At the heart of NBEs are neural distribution functions: a physics-inspired parameterisation that can be optimised with neural networks conditioned on theory parameters, enabling efficient parameter scans. The natural gradient method allows the training of conditional and unconditional distribution functions to high precision.

NBE importance sampling is $1000\times$ more sample-efficient than a recently proposed Monte Carlo Boltzmann solver, which allowed us to evolve a complex non-equilibrium system with number-changing $2\leftrightarrow 3$ processes to per-mil precision. We then performed a precision calculation of the effective number of relativistic neutrino degrees of freedom with a $50\times$ smaller solver uncertainty, setting the stage for future calculations including additional higher-order effects. Finally, we used a single conditioned NBE for a two-dimensional parameter scan of an extended $N_\mathrm{eff}$ scenario motivated by the current tension between experiment and the Standard Model.

Together, these results establish NBEs as a precise, flexible, and scalable framework for early-universe calculations, bringing higher-multiplicity processes and high-dimensional parameter spaces within reach of precision early-universe calculations.

\newpage
\subsubsection*{AI use statement}

In this work, we used generative AI tools to help develop theoretical models or conceptual frameworks, propose or refine hypotheses, design or provide feedback on research methodology or experiments, to implement methods, and to interpret results. We have not used generative AI tools to formulate mathematical claims, provide critical ingredients for proving mathematical claims, assist in the writing of proofs, assist in translation. The use cases of generating synthetic datasets, cleaning and reformatting datasets, and supporting qualitative and thematic data analysis are not applicable to this work. Additionally, we used generative AI tools to create or modify scientific figures or images, to create or edit software code, to summarise or analyse existing literature, and for brainstorming. We have reviewed all AI-assisted work. We used AI extensively, both to get feedback on ideas and to turn those ideas into efficient implementations. Because AI is increasingly good at assembling brute-force solutions, i.e. combining many small ingredients until a task works, without making clear which of them actually matters, we focused on distilling the minimal set of ingredients the method needs. We verified and tested AI-generated code by working through it and rewriting it to improve readability. We take responsibility for the final content of this work, including text, claims or artefacts produced with the aid of generative AI.

\subsubsection*{Reproducibility statement}
We describe the method in more detail in Appendix~\ref{app:method}, and give extended information about our experiments in Appendix~\ref{app:experiment-details}. We provide the analytic expressions for the neutrino decoupling calculation in Appendix~\ref{app:neff-theory} for completeness, even though they are also available in the literature.
We attached our implementation in the supplementary materials and will publish it upon acceptance. It contains a file \texttt{REPRODUCE.md} with all commands required to obtain the main results, experiment configurations in the \texttt{config/} folder, and the folder \texttt{paper/} contains the code to reproduce all figures based on the runs. No external datasets are required.

\subsubsection*{Acknowledgments}
We thank Marco Drewes, Yuan-Zhen Li, Sergio Pastor, Maksym Ovchynnikov, and Miguel Escudero Abenza for valuable discussions, and Theo Heimel and Ramon Winterhalder for help with MadSpace and MadNIS. 
JDS gratefully acknowledges support from the UK Research and Innovation Future Leader Fellowship MR/Y018656/1.
JS gratefully acknowledges support from the Alexander von Humboldt foundation as a Feodor Lynen Fellow.

\bibliography{literature}
\bibliographystyle{iclr2027_conference}

\clearpage
\appendix
\input{appendix.tex}

\end{document}

%% file: math_commands.tex
\usepackage{amsmath,amsfonts,bm}

\def\eqref#1{equation~\ref{#1}}

\def\1{\bm{1}}

\DeclareMathAlphabet{\mathsfit}{\encodingdefault}{\sfdefault}{m}{sl}
\SetMathAlphabet{\mathsfit}{bold}{\encodingdefault}{\sfdefault}{bx}{n}



%% file: appendix.tex
\section{Method details}
\label{app:method}

\subsection{Neural distribution functions}
\label{app:distribution-function}

\textbf{Degrees of freedom.}
The formalism in Section~\ref{sec:neural_distribution_functions} focuses on the case of a particle with a single degree of freedom for simplicity. However, in the neutrino decoupling calculation in Section~\ref{sec:Neff} we describe the electron species with four degrees of freedom. As we begin our evolution without asymmetries, and there are no effects that distinguish between particle and antiparticle, or between left and right helicity, all share the same distribution function. This is accounted for by factors of the relativistic degrees of freedom for each particle, $g$, in \eqref{eq:distribution-function},
\begin{align}
    f(y) = \frac{4\pi^2 E}{gy^2}\eta p(y)\;,\qquad \eta = g\int \dd y\frac{y^2}{4\pi^2 E}f(y)\;.
\end{align}
Electrons, for example, have $g=4$. The moments of the distribution, defined in Section~\ref{sec:background}, become $M_a[\phi] = g_a\int \dd^3 y/(2\pi)^3 \phi (y)f_a(y)$, and are the summed contributions from each individual degree of freedom. For the collision integral, we work in terms of the spin-summed matrix elements $|\mathcal{M}_k|^2$ which leaves the one-particle Lorentz-invariant phase-space measure $\dd\Pi_a$ unchanged, as it still refers to a single degree of freedom.

\textbf{Equilibrium solutions with neural distribution functions.}
The ansatz of \eqref{eq:gamma-mixture} using the kinetic energy $\varepsilon=(E-m)/T$ is chosen such that equilibrium distributions can naturally be represented. Expanding the Bose-Einstein and Fermi-Dirac distributions in a geometric series, we find
\begin{align}\label{eq:feq-series}
    f_\mathrm{eq}(y) = \frac{1}{e^{(E-\mu)/T}\pm 1} = \sum_{k=1}^\infty (\mp 1)^{k+1} e^{-k(m-\mu)/T} e^{-k\varepsilon}\;,\quad a=\frac{m}{T}\;,\quad \varepsilon = \frac{E-m}{T}\;.
\end{align}
Inserting this into \eqref{eq:distribution-function} with $y=T\sqrt{\varepsilon(\varepsilon+2a)}$, $y^2\,\dd y/E=Ty\,\dd\varepsilon$ gives
\begin{align}\label{eq:peq-series}
p_\text{eq}(y)\, \dd y&=\frac{T^2}{4\pi^2\eta_\text{eq}}\sum_{k=1}^{\infty}(\mp 1)^{k+1}e^{-k(m-\mu)/T}\,
\sqrt{\varepsilon(\varepsilon+2a)}\;e^{-k\varepsilon}\,\dd\varepsilon\;,\notag\\
\eta_\text{eq}&=\frac{mT}{4\pi^2}\sum_{k=1}^{\infty}(\mp 1)^{k+1}\frac{e^{k\mu/T}}{k}K_1(ka)\;.
\end{align}
The fermionic series has alternating signs and can therefore not be parameterised as a gamma mixture distribution, motivating the $f=f_0/(1+f_0)$ parameterisation. The density $p_\mathrm{eq}(y)$ reduces to a series of gamma mixture distributions for massless particles, $a=0$, and in the non-relativistic limit, $a\to\infty$. For a massless boson with $\mu=0$, we find
\begin{equation}
p_\text{eq}(y)\,\dd y=\frac{6}{\pi^2}\sum_{k=1}^{\infty}\frac{1}{k^2}\,\gamma(\varepsilon|2,k)\,\dd\varepsilon\;,
\qquad \eta_\text{eq}=\frac{T^2}{24}\;.
\end{equation}
In the non-relativistic limit, only the $k=1$ term survives and we obtain the Maxwell-Boltzmann distribution, which is a single gamma distribution, $p_\mathrm{eq}(y)\,\dd y=\gamma (\varepsilon | 3/2, 1)\dd\varepsilon$. For general masses, we interpolate between the two limits,
\begin{equation}
    \sqrt{\varepsilon(\varepsilon+2a)}=\big(\varepsilon+\sqrt{2a\varepsilon}\big)\,A(\varepsilon)\;, \qquad A(\varepsilon)=\frac{\sqrt{\varepsilon+2a}}{\sqrt{\varepsilon}+\sqrt{2a}}\in\left[\frac{1}{\sqrt{2}}, 1\right]\;.
\end{equation}
The exact equilibrium density is therefore a mixture of $\gamma (\varepsilon | 2,k)$ and $\gamma (\varepsilon | 3/2,k)$ components, with a bounded correction factor $A(\varepsilon)$. We use this form to sample exact equilibrium distributions by rejection. A learned distribution function instead absorbs the $A(\varepsilon )$ prefactor into the mixture parameters $(\alpha_m, \lambda_m)$.

For fermions, \eqref{eq:fermions-distribution-function} maps the Fermi-Dirac distribution onto the Maxwell-Boltzmann distribution as $f_0(y)=e^{-(E-\mu)/T}$. The corresponding rejection step accepts a proposal $y\sim p_0(y)$ with probability $1/(1+f_0(y))$ and efficiency $N$. For the massless Fermi-Dirac distribution at $\mu=0$, this gives $N=\pi^2/12\approx 0.8$.

\textbf{Skewed gamma mixture distributions.}
The design choice $p(y)\propto y^2 f/E$ is based on the invariant measure $\dd^3 y/E$, which makes it the natural proposal distribution for the initial states of the collision integral. However, the fit function $g=y^2f$ corresponds to the integration measure $\dd^3 y$, for which $p(y)$ is not the optimal proposal distribution. To overcome this, we define the skewed density
\begin{equation}
    p_\ell (y)= \frac{E^\ell}{\mathbb{E}_{y\sim p(y)}[E^\ell]} p(y)\;.
\end{equation}
Using the skewed density $p_1$ to sample $y\sim p_1(\cdot |\omega_{n+1,a})$ in Algorithm~\ref{alg:time-step} improves the sampling efficiency compared to the naive choice $p(\cdot |\omega_{n+1,a})$. We recover the original density with $p_0=p$, and use $p_1$ to sample the test momenta in Alg.~\ref{alg:time-step}. The skewed density $p_\ell$ is again a gamma mixture distribution: using $E=m+T\varepsilon$, we find for $\ell =1$
\begin{equation}
    E \gamma (\varepsilon | \alpha, \lambda ) = m\gamma (\varepsilon | \alpha, \lambda ) + T\frac{\Gamma (\alpha+1)}{\Gamma (\alpha)\lambda} \gamma (\varepsilon | \alpha +1, \lambda )\;.
\end{equation}
In this way, we obtain from a gamma mixture $p$ another gamma mixture $\propto Ep$ by doubling the number of mixture components and updating the normalisation. Expressions for $\ell >1$ can be derived in the same way, but require increasingly many mixture components. Note that this trick only works for $\ell\ge 0$, which is the reason why we define the parameterisation based on the $\dd^3 y/E$ measure. For fermions, we draw from the skewed variant $p_{0,1}$ of the bosonic density $p_0$ and apply the standard rejection step with probability $1/(1+f_0)$. We find that this modification improves the efficiency of the Monte Carlo estimate, and use it in all experiments.

\textbf{Moments of neural distribution functions.}
With the factorisation of \eqref{eq:distribution-function}, any moment of $f$ reduces to an expectation under the normalised density,
\begin{equation}
\langle h\rangle_f\equiv\int\frac{\dd^3y}{(2\pi)^3}\,h(y)\,f(y) =\frac{1}{2\pi^2}\int \dd y\,y^2\,h(y)\,\frac{4\pi^2E}{y^2}\,\eta\, p(y) =2\eta\;\mathbb{E}_{y\sim p(y)}\big[E\,h(y)\big]\;.
\end{equation}

For fermions, \eqref{eq:fermions-distribution-function} gives $p=p_0/(N(1+f_0))$ and $\eta=N\eta_0$, so the normalisation drops out, and we find
\begin{equation}
    \langle h\rangle_f=2\eta_0\,\mathbb{E}_{y\sim p_0(y)}[E\,h/(1+f_0)]\;.
\end{equation}
The gamma mixture is a density in $\varepsilon$, and $E=m+T\varepsilon$, $y^2=T^2\varepsilon(\varepsilon+2a)$ are polynomial in $\varepsilon$, so every moment with polynomial $h$ follows from
\begin{equation}
    \mathbb{E}_{y\sim p(y)}\left[\varepsilon (y)^q\right] = \sum_{m=1}^M w_m \int \dd \varepsilon\, \varepsilon^q \gamma (\varepsilon|\alpha_m, \lambda_m ) = \sum_{m=1}^M w_m \frac{1}{\lambda_m^q}\frac{\Gamma (\alpha_m+q)}{\Gamma (\alpha_m)}\;.
\end{equation}
For general $h$, and for all fermion moments because of the factor $1/(1+f_0)$, we use Gauss-Laguerre quadrature with $N_\text{quad}$ nodes $t_i$ and weights $v_i$, $\int_0^\infty \dd t\,e^{-t}g(t)\approx\sum_i v_i\,g(t_i)$. Substituting $r=\lambda_m\varepsilon$ in each component,
\begin{align}
    \mathbb{E}_{y\sim p(y)}\left[h(y)\right] &= \sum_{m=1}^M w_m \frac{\lambda_m^{\alpha_m}}{\Gamma (\alpha_m)} \int \dd \varepsilon\, h(\varepsilon) \varepsilon^{\alpha_m-1}e^{-\lambda_m \varepsilon}\;,\notag\\
    &= \sum_{m=1}^M \frac{w_m}{\Gamma (\alpha_m)} \int \dd r\, h(r/\lambda_m) r^{\alpha_m-1} e^{-r}\;,\notag\\
    &= \sum_{m=1}^M \frac{w_m}{\Gamma (\alpha_m)}\sum_{i=1}^{N_\mathrm{quad}} v_i h(t_i/\lambda_m) t_i^{\alpha_m-1}\;.
\end{align}

For massive particles the kinematic factor $y=T\sqrt{\varepsilon(\varepsilon+2a)}$ has a $\sqrt{\varepsilon}$ branch point at $\varepsilon=0$ that spoils the spectral convergence of the Laguerre rule. Substituting $t=u^2$ maps it to a Gaussian integral over the half line, $\int_0^\infty \dd t\,e^{-t}g(t)=\int_0^\infty 2u\,\dd u\,e^{-u^2}g(u^2)$, where $\sqrt t=u$ is polynomial. We evaluate it with the positive nodes $u_i$ of a $2N_\text{quad}$-point Gauss-Hermite rule, i.e. we use the rule above with $t_i=u_i^2$ and $v_i=2u_iv_i^{\rm H}$. We switch from Laguerre to Hermite nodes for $a=m/T>0.1$.

\subsection{Collision integral}
\label{app:collision}

Here we give additional information on the collision integral evaluation described in Section~\ref{sec:collision_integrals}, and in particular, in Algorithm~\ref{alg:collision-sampling}. We use the \texttt{TPropagatorMapping} of \citet{Heimel:2026hgp} with coefficient $0.7$ as the phase-space mapping, denoted $\mathrm{MadSpace}$ in Algorithm~\ref{alg:collision-sampling}.

\textbf{Distribution function sampling.}
The factorisation of $f(y)$ into the invariant density, $\eta$, and the normalised density, $p(y)$, in \eqref{eq:distribution-function} is designed to enable efficient importance sampling for the collision integral initial states. Sampling the angles uniformly $\hat\Omega\sim \mathcal{U}(S^2)$, and momenta as $y\sim p(y)$, we obtain for the corresponding phase-space measure
\begin{equation}
    \frac{\dd\Pi}{(4\pi)^{-1}p(y)} = \frac{\dd\Omega}{(4\pi)^{-1}} \frac{y^2\dd y}{(2\pi)^3 2E} \frac{1}{p(y)} = \frac{\dd\Omega}{(4\pi)^{-1}} \frac{y^2 \dd y}{(2\pi)^3 2E} \frac{4\pi^2 E}{y^2} \frac{\eta}{f(y)} = \dd\Omega\,\dd y\frac{\eta}{f(y)}\;.
\end{equation}
The factors in $p(y)$ cancel with those in the Lorentz-invariant phase-space measure, $\dd\Pi$, by construction, leaving a factor $\eta/f(y)$, with the $f(y)$ factor naturally cancelling with the factors of $f(y)$ or $1\pm f(y)$ in the statistical factor $\Lambda[f]$. For neural importance sampling the initially sampled momentum, $\tilde y$, is modified with the flow, $\tilde y\to y$, and we then obtain $f(y)$ as $f(y)=(4\pi^2 E)/y^2 p(\tilde y)$, where the prefactor depends on the modified $y$ and the initially sampled $\tilde y$ appears in the density.

\textbf{Neyman allocation.}
When multiple processes contribute to a collision integral, using the same number of Monte Carlo samples $N_i$ for each process is typically inefficient, because the total variance is dominated by the process with the largest per-sample variance. Instead, we dynamically set the sample count $N_i$ based on the total process standard deviations, $\sigma_i$, of the previous time step. For a given sample budget, $N_\mathrm{tot}$, we use the so-called Neyman allocation
\begin{equation}
    N_i = N_\mathrm{tot}\frac{\sigma_i}{\sum_j \sigma_j}\;.
\end{equation}

\textbf{Phase-space mappings.} 
Phase-space mappings turn uniform random numbers into kinematic configurations that satisfy on-shell relations and energy-momentum conservation. The mapping is constructed such that the corresponding Jacobian transforms the uniform base density into a phase-space density that is a good approximation to the integrand, minimising integral variance, see \eqref{eq:mc-integration}. A range of phase-space mappings exist for different structures in the integrand, such as the Rambo mapping~\citep{Kleiss:1985gy} for an integrand that is constant up to the phase-space measure $\dd\Phi_k$, or a Breit-Wigner mapping for a resonant structure in the matrix element. Modern event generators build custom phase-space mappings recursively based on Feynman diagram topologies, see \citet{Heimel:2026hgp} for a recent review. Our implementation uses the MadSpace package~\citep{Heimel:2026hgp} which supports all the options discussed above.

\textbf{$\mathbf{ToUnitCube}$ mapping.}
The $\mathrm{ToUnitCube}$ mapping in Alg.~\ref{alg:collision-sampling} maps the momenta $y_i\in\mathbb{R}$, angles $\phi_i\in [0,2\pi]$, $\cos\theta_i\in [-1,1]$, and $u\in [0,1]$ to the unit hypercube $[0,1]^N$. The optimal choice is to use the corresponding cumulative distribution function, which ensures that the resulting unit hypercube is populated uniformly. The $u$ parameters need no modification and for the uniformly sampled angles, $\phi_i$ and $\cos\theta_i$, a shift and rescaling is sufficient. For the momenta $y$, we use a scaled exponential transform as a simple proxy for the full gamma mixture distribution, for which the cumulative distribution function is not available in closed form. Using the same mapping for each species and writing $\hat\Omega_i = (\phi_i, \cos\theta_i)$, we arrive at the mapping
\begin{equation}
    \mathrm{ToUnitCube}(y_2, \dots \hat\Omega_1,\dots u) = \left(1-\exp\left(-\frac{\varepsilon_2}{\mathbb{E}[\varepsilon_2]}\right),\dots \left(\frac{\phi_1}{2\pi}, \frac{1+\cos\theta_1}{2}\right),\dots ,u\right)\;.
\end{equation}

\subsection{Boltzmann Equation training}
\label{app:training}

\textbf{Hubble parameter.}
The Hubble parameter is given in terms of the Planck mass $m_\mathrm{Pl}$ and the total energy density,
\begin{equation}
    H = \sqrt{\frac{8\pi \rho_\mathrm{tot}}{3m_\mathrm{Pl}^2}}\;.
\end{equation}
Note that $m_\mathrm{Pl}$ is the Planck mass in comoving units, related to the physical Planck mass by $m_\mathrm{Pl} = (x/m_0) m_\mathrm{Pl}^\mathrm{phys}$.

\textbf{Integrator schemes.}
Here we specify the discretisation used for the time evolution in terms of a generic scheme $(A,\gamma)$ in \eqref{eq:simple-boltzmann}. We implement three specific schemes:
\begin{align}\label{eq:schemes}
    \text{Euler}:\qquad &A=g_n,\qquad \gamma=1,\notag\\
    \text{BDF2}:\qquad &A=ag_n + bg_{n-1},\qquad \gamma = \frac{1+r}{1+2r},\notag\\
    &a=\frac{(1+r)^2}{1+2r},\qquad b=-\frac{r^2}{1+2r},\qquad r=\frac{\log x_{n+1}-\log x_n}{\log x_n-\log x_{n-1}}\;,\notag\\
    \text{CN}:\qquad &A=g_n + \frac{c_a g_n + c_b g_{n-1}+c_c g_{n-2}}{2},\qquad \gamma=\frac{1}{2},\notag\\
    &c_a=\frac{r_1 (2r_2+1)}{1+r_2}, \qquad c_b=-r_1(1+r_2),\qquad c_c=\frac{r_1r_2^2}{1+r_2},\notag\\ &r_1=\frac{\log x_{n+1}-\log x_n}{\log x_n-\log x_{n-1}}, \qquad r_2=\frac{\log x_n-\log x_{n-1}}{\log x_{n-1}-\log x_{n-2}}\;.
\end{align}
The Euler scheme is the standard single-step approach. The BDF2 scheme combines information from two previous time steps. Our most powerful and default choice, the CN scheme, starts from $A=g_n+\gamma h s(g_n)$, then improves the collision integral estimate with the trapezoid rule as $s(g_n)\gets (s(g_n)+s(g_{n+1}))/2$, and then inserts an estimate for $s(g_n)$ based on the previous solutions. In each case, the coefficients above are obtained from solving a linear system of conditions that match derivatives at the previous solutions. We find that the CN and BDF2 schemes significantly improve the solver precision.

\textbf{Collision operator derivative estimate.}
The iteration speed when solving $g_{n+1}'=T(g_{n+1}) = g_{n+1}-M^{-1}F(g_{n+1})$ of Section~\ref{sec:training} depends on the choice of $M$. The ideal choice $M=\partial F/\partial g$, following from the optimal update to leading order $\partial T/\partial g=1-M^{-1}\partial F/\partial g=0$, is impractical due to the large time and memory cost of differentiating the collision operator. On the other hand, the naive choice $M=1$, corresponding to $\partial s/\partial g=0$, suffers from unstable numerical behaviour. To find an optimal middle ground, we approximate the collision operator derivative of species $a$:
\begin{align}\label{eq:kappa-expression}
    \frac{\partial s_a}{\partial g_a} &= \frac{\partial}{\partial f_a} \frac{1}{2E_aH} \sum_k \int \dd\Phi_{k\setminus a} |\mathcal{M}_k|^2 \Lambda_k[f]\;,\notag\\
    &= \frac{\partial}{\partial f_a}\frac{1}{2E_aH}\sum_k \int \dd\Phi_{k\setminus a} |\mathcal{M}_k|^2 \left((1\mp f_a)\prod_{b\in k_\mathrm{out}}f_b\prod_{b\in k_\mathrm{in}\backslash 1}(1\mp f_b) - f_a(\mathrm{in}\leftrightarrow\mathrm{out})\right)\;,\notag\\
    &= -\frac{\partial}{\partial f_a}\left[\frac{1}{2E_aH}\sum_k \int\dd\Phi_{k\setminus a} |\mathcal{M}_k|^2f_a\left(\pm \prod_{b\in k_\mathrm{out}}f_b\prod_{b\in k_\mathrm{in}\backslash 1}(1\mp f_b) + (\mathrm{in}\leftrightarrow\mathrm{out})\right)+\mathrm{const}\right]\;,\notag\\
    &\approx -\frac{1}{2E_aH}\sum_k \int\dd\Phi_{k\setminus a} |\mathcal{M}_k|^2 \left(\pm \prod_{b\in k_\mathrm{out}}f_b\prod_{b\in k_\mathrm{in}\backslash 1}(1\mp f_b) + (\mathrm{in}\leftrightarrow\mathrm{out})\right)\equiv -\kappa_a\;.
\end{align}
In the last step, we have neglected $f_a$ occurrences in the matrix element and on other legs of the collision operator. The resulting approximation to $\partial s(y)/\partial g(y')$ is exact on the diagonal, $y=y'$. For bosons, stimulated emission allows $\kappa <0$, which can lead to unstable regimes. We set $\kappa = \max (\kappa, 0)$ to avoid this.

\textbf{Natural gradient update with conserved moments.}
If certain moments of a species such as the number density $n$ or energy density $\rho$ are conserved, then the corresponding direction in the neural distribution function parameter landscape can suffer from systematic shifts from the natural gradient step that build up over time steps. This does not happen for non-conserved moments, because collisions damp the effect. To correct for this, we first construct a natural gradient update that acts on the parameter directions that leave conserved moments unchanged, and afterwards do a low-dimensional refinement update on the conserved moments.

We consider a list of weighting functions $\psi = (\psi_1,\dots ,\psi_n)$ with moments $\Psi[g]=\int \dd y\,\psi (y)g(y)$ that are conserved by collisions, $\Psi [s(g)]=0$. We use $\psi =(1,E)$ throughout this work to handle number and energy density conservation. Including moments in $\Psi$ that are not conserved does not reduce the precision of the update, but marginally increases the cost.

To remove this shift, we want to enforce $\Delta\Psi=\Psi [g-g_\mathrm{target}]=0$ in addition to the standard objective of \eqref{eq:natgrad-setup}. We estimate the moments $\Psi [g]$ with Monte Carlo integration, $\Psi [g] = \mathbb{E}_{y\sim p(y)}[\psi (y) g(y) / p(y)]$. Defining the corresponding Jacobian $D=\partial\Delta\Psi /\partial\theta$, we obtain $D\delta\theta = -\Delta\Psi$ as a constraint on the parameter update $\delta\theta=\theta_*-\theta_0$ in \eqref{eq:natgrad-step}. We extend the objective $L(\theta )$ of \eqref{eq:natgrad-setup} with a Lagrange multiplier $\lambda$ to enforce this condition, and obtain
\begin{equation}
    \begin{pmatrix}F & D^\top \\ D & 0\end{pmatrix} \begin{pmatrix}\delta\theta\\\lambda\end{pmatrix} = \begin{pmatrix}-J^\top Wr\\-\Delta\Psi\end{pmatrix}\;.
\end{equation}
Inserting the first row, $\delta\theta=\delta\theta_\mathrm{NG}-F^{-1}D^\top\lambda$ with $\delta\theta_\mathrm{NG}=-F^{-1}J^\top Wr$, into the second row yields the solution $\lambda = S^{-1}(D\delta\theta_\mathrm{NG}+\Delta\Psi )$ for the Lagrange multiplier, with the matrix $S=DF^{-1}D^\top$. Substituting this back into the first row yields the solution for the constrained natural gradient step
\begin{equation}
    \delta\theta = \left(\mathbb{I} - F^{-1}D^\top S^{-1}D\right)\delta\theta_\mathrm{NG} - F^{-1}D^\top S^{-1}\Delta\Psi\equiv \delta\theta_\text{shape} + \delta\theta_\text{moment}\;.
\end{equation}
The shape step $\delta\theta_\text{shape}$ leaves the moments unchanged, $D\delta\theta_\text{shape}=0$, and the moment step $\delta\theta_\text{moment}$ shifts the moments by the required amount, $D\delta\theta_\text{moment}=-\Delta\Psi$, using a minimal step in parameter space as defined by the Fisher metric. The two steps commute only at linear order in $\delta\theta$, and we therefore first apply the shape step $\delta\theta_\text{shape}$ and do the moment step $\delta\theta_\text{moment}$ afterwards to ensure that the moment change is correct. In practice, we first determine $D$ and $S$ in addition to $F$, construct $F^{-1}$ and $S^{-1}$ with the usual regularised matrix inversion, then construct and apply the shape step $\delta\theta_\text{shape}$, and finally perform the moment step $\delta\theta_\text{moment}$ by iteratively refining the parameters and evaluating $\Delta\Psi$ at updated parameters until the moments match.

See Appendix~\ref{app:moment-pins} for a study on the effect of this extension of the natural gradient update. We find that this constraint removes systematic shifts in conserved moments, and therefore use it in all our experiments. The computational cost is negligible compared to the collision integral evaluation.

\textbf{Time step convergence criteria.}
In each step of Algorithm~\ref{alg:time-step}, we test the following two convergence criteria after constructing the residual $r$:
\begin{enumerate}
    \item Noise floor: The residual is dominated by Monte Carlo noise, $\mathrm{median}_y[r^2(y) / \sigma_r^2(y)] \le c \chi^2_{1,\mathrm{median}}$, where $\chi^2_{1,\mathrm{median}}\approx 0.455$ is the median of the $\chi^2_1$ distribution and we use $c=1.2$.
    \item Plateau: The objective $r^\top Wr$ decreased by less than a factor $r_\mathrm{plateau}$ over the last three iterations with $r_\mathrm{plateau}=0.05$, indicating that the residual is limited by the expressivity of the ansatz or the solver.
\end{enumerate}
If either criterion is satisfied, we accept the current parameters $\theta_{n+1}$ instead of performing another natural gradient update. This typically occurs after 5-10 iterations for unconditional trainings. For conditional trainings, we additionally require a minimum number of iterations before accepting convergence, typically 15, to ensure that the network sees a sufficient range of conditions. Larger condition spaces and stronger dependences on the condition require a larger batch size or more startup iterations.

\textbf{Step size controller.}
We use a dynamic step size controller~\citep{Gustafsson:1991,Soderlind:2003} to determine the step size $h_n = \log x_{n+1}-\log x_n$ in \eqref{eq:simple-boltzmann}, based on the deviation of the fixed-point target from the linear extrapolation of the previous solutions, $e(y) = g_\mathrm{target}(y)-g_n(y)-r_n(g_n(y)-g_{n-1}(y))$, $r_n=h_n/h_{n-1}$. We use this error to construct the error ratio $\zeta_n$ which is used to compute the next step size, $h_{n+1}$,
\begin{align}
    \zeta_n &= \max_\mathrm{species}\mathbb{E}_{y\sim p(y)}\left[\left(\frac{e(y)}{\epsilon_\mathrm{abs}+\epsilon_\mathrm{rel}|g_n(y)|+c\sigma_g (y)}\right)^2\right]^{1/2}\;,\notag\\
    h_{n+1} &= h_n\beta \zeta_n^{-k_I/2}\left(\frac{\zeta_{n-1}}{\zeta_n}\right)^{k_P/2}\;.
\end{align}
We use the values $\beta =0.9$, $k_I=0.7$, $k_P=0.4$. The Monte Carlo uncertainty of the target $g_\mathrm{target}(y)$ is given by $\sigma_g (y)=\gamma h_n \sigma_s(y) / (1+\gamma h_n\kappa)$, where $\sigma_s(y)$ is the Monte Carlo uncertainty of the collision integral estimate $s(y)$. The $\sigma_g$ term in $\zeta_n$ implements a noise floor that prevents step size decrease if $e(y)$ is dominated by noise. We set $c=2$ but find that this parameter has little impact in our experiments.
The absolute and relative tolerances $\epsilon_\mathrm{abs}$ and $\epsilon_\mathrm{rel}$ set the tolerance band for $e(y)$, typically $\epsilon_\mathrm{abs}\sim 10^{-8}$, $\epsilon_\mathrm{rel}\sim 10^{-6}$.
We further enforce $h_{n+1}\in [\Delta\log x_\mathrm{min},\Delta\log x_\mathrm{max}]$ to prevent problems from very large or small steps.

\section{Neutrino decoupling formalism}
\label{app:neff-theory}

Standard Model neutrino decoupling serves as a precision test of Boltzmann equation solvers. We use the fully momentum-dependent, three-flavour transport treatments introduced in several previous works~\citep{Mangano:2005cc, deSalas:2016ztq, Akita:2020szl, Froustey:2020mcq, Bennett:2020zkv}, and assume FLRW spacetime, no lepton asymmetry, no CP-violating phase in the lepton sector, massless neutrinos for the collision kinematics, and truncate weak collision rates at leading order in the Fermi constant, $G_F$. We include finite-temperature quantum electrodynamic (QED) corrections to the equation of state up to $\mathcal{O}(e^5)$, and assume that photons and electrons remain in exact equilibrium at all times. The comoving variables introduced in Section~\ref{sec:background} are used throughout.

\textbf{Quantum kinetic equation.} Accounting for the effects of neutrino mixing promotes the neutrino distribution functions, $f_{\alpha}$, $\alpha \in \{e,\mu,\tau\}$, to the Hermitian density matrix $F$, whose components are defined by the two-point correlation function
\begin{equation}
    \langle a^\dagger_\beta(\vec{y}) a_\alpha(\vec{y}')\rangle = (2\pi)^3 2E_y \delta^{(3)}(\vec{y} - \vec{y}') F_{\alpha\beta}(y)\;.
\end{equation}
The diagonal elements of $F$ are the occupations of each flavour eigenstate, whilst the off-diagonal components encode coherence. The antiparticle density matrix, $\bar{F}$, has components $\bar F_{\alpha\beta}(y) \sim \langle b^\dagger_\alpha b_\beta\rangle$, and is related to the particle density matrix by $\bar F = F^\top$ in the absence of CP violation and lepton asymmetry. 

The density matrix evolves according to the quantum kinetic equation (QKE)~\citep{Sigl:1993ctk}
\begin{equation}\label{eq:QKE}
    Hx\frac{\partial F}{\partial x} = -i[\mathcal{H}_\mathrm{eff},F] + \mathcal{C}_\nu[F], \quad \mathcal{H}_\mathrm{eff} = \frac{U \Delta M^2 U^\dagger}{2y} - \frac{4 G_F y}{\sqrt{2} m_W^2}(\rho_e + P_e) \Pi_e - \frac{8\sqrt{2}G_F y}{3 m_Z^2} \varrho_\nu\;,
\end{equation}
where $U$ is the Pontecorvo–Maki–Nakagawa–Sakata neutrino mixing matrix~\citep{Pontecorvo:1957qd,Maki:1962mu}, $\Delta M^2 = \operatorname{diag}(0,\Delta m_{21}^2, \Delta m_{31}^2)$, $\Pi_e = \operatorname{diag}(1,0,0)$, and $m_W$ and $m_Z$ are the weak gauge boson masses.
The first term in $\mathcal{H}_\mathrm{eff}$ governs neutrino oscillations, transferring occupation between flavour eigenstates. The remaining terms describe coherent forward scattering in the early-universe plasma, which modify the oscillation frequencies. The $\varrho_\nu$ term is proportional to the identity, and so vanishes under the commutator, when the small differences between neutrino flavours are neglected~\citep{Froustey:2020mcq}. As such, we will not include it going forward. Neither term modifies the total occupation, captured by $\operatorname{Tr}[F]$. 

The energy density, $\rho_a$, and pressure, $P_a$, of a species are moments of the distribution functions
\begin{equation}\label{eq:moments}
    \rho_a = g_a\int \frac{\dd^3y}{(2\pi)^3} E_a f_a, \qquad P_a = g_a\int \frac{\dd^3y}{(2\pi)^3} \frac{y^2}{3E_a} f_a, \qquad \varrho_\nu = 2\int \frac{\dd^3y}{(2\pi)^3} y \operatorname{Re}[F]\;, 
\end{equation}
with $g_e = 4$ for electrons, and $g_\gamma = 2$ for photons. The full matrix $\varrho_\nu$ contributes to coherent forward scattering, but only its trace, $\rho_\nu = \operatorname{Tr}[\varrho_\nu]$, proportional to the physical neutrino energy density, contributes to gravitational interactions. If we neglect neutrino mixing, $F$ is diagonal and the commutator $[\mathcal{H}_\mathrm{eff},F]$ vanishes, recovering the Boltzmann equation of~\eqref{eq:BoltzmannEq}. 

\textbf{Collision operator.} For a process $r = a(y_1) + b(y_2) \leftrightarrow c(y_3) + d(y_4)$, the QKE collision operator is the matrix-valued expression
\begin{align}
    \mathcal{C}^{(a)}_\nu[F] = &\frac{1}{2E_1} \sum_r S^{(r)} \int \left(\prod_{i=2}^4 \dd\Pi_i\right) (2\pi)^4 \delta^{(4)}(y_1 + y_2 - y_3 -y_4) \notag\\
    &\times\left[\frac{1}{2}\left\{\mathcal{K}_\mathcal{G}^{(r)},\widetilde F^{(a)}_1\right\} - \frac{1}{2}\left\{\mathcal{K}_\mathcal{L}^{(r)}\;, F^{(a)}_1\right\}\right], \qquad \widetilde F = \mathbb I -F\;,
\end{align}
with $S^{(r)}$ the symmetry factor. The superscript $(a)$ denotes the particle species, $(r)$ the process, and the subscript the momentum label, \textit{e.g.} $F_1^{(a)} = F^{(a)}(y_1)$. The gain and loss kernels sum over all spin and flavour interaction channels:
\begin{equation}
    \left(\mathcal{K}_\mathcal{G}^{(r)}\right)_{\alpha\beta} = \sum_{\mathrm{spins}} \sum_{\mu,\mu'}\sum_{\nu,\nu'}\sum_{\rho,\rho'} \mathcal{M}_{\alpha\mu;\nu\rho}^{(ab;cd)*} \mathcal{M}_{\beta\mu';\nu'\rho'}^{(ab;cd)} \left(\widetilde F_2^{(b)}\right)_{\mu'\mu} \left( F_3^{(c)}\right)_{\nu\nu'} \left( F_4^{(d)}\right)_{\rho\rho'}\;,
\end{equation}
where $\mathcal{M}_{\alpha\mu;\nu\rho}^{(ab;cd)}$ is the amplitude for $a_\alpha b_\mu\to c_\nu d_\rho$ with momenta assigned as above. The loss kernel follows from the replacements $F \leftrightarrow \widetilde F$, and $f \leftrightarrow \widetilde{f}$ for scalar distributions.

There are five processes contributing to neutrino decoupling, with gain kernels~\citep{Froustey:2020mcq,Bennett:2020zkv}
\begin{align}
    \mathcal{K}_\mathcal{G}^{(\nu e;\nu e)} &= 32 G_F^2\widetilde{f}_2^{(e)} f_4^{(e)} W_{\nu e}\left[F^{(\nu)}_3\right]\;,\notag\\
    \mathcal{K}_\mathcal{G}^{(\nu \bar e;\nu \bar e)} &= 32 G_F^2\widetilde{f}_2^{(e)} f_4^{(e)} W_{\nu \bar e}\left[F^{(\nu)}_3\right]\;, \notag\\ 
    \mathcal{K}_\mathcal{G}^{(\nu \bar \nu;e \bar e)} &= 32 G_F^2f_3^{(e)} f_4^{(e)} W_{\nu \bar \nu}\left[\left(\widetilde{F}^{(\nu)}_2\right)^\top\right]\;, \notag\\  
    \mathcal{K}_\mathcal{G}^{(\nu \nu;\nu \nu)} &= 8G_F^2 s^2 \left(F_3^{(\nu)} \mathcal{T}\left[\widetilde{F}^{(\nu)}_2 F^{(\nu)}_4\right] + F_4^{(\nu)} \mathcal{T}\left[\widetilde{F}^{(\nu)}_2 F^{(\nu)}_3\right]\right)\;,\notag\\
    \mathcal{K}_\mathcal{G}^{(\nu \bar\nu;\nu \bar\nu)} &= 8G_F^2 u^2 \left(\mathcal{T}\left[\left(\widetilde{F}^{(\nu)}_2\right)^\top \left(F^{(\nu)}_4\right)^\top\right]F_3^{(\nu)}  +\mathcal{T}\left[F^{(\nu)}_3 \left(F^{(\nu)}_4\right)^\top\right] \left(\widetilde F_2^{(\nu)}\right)^\top\right)\;,
\end{align}
in terms of the Mandelstam variables $s = (y_1+y_2)^2$, $t = (y_1-y_3)^2$, and $u = (y_1 - y_4)^2$, and matrix maps
\begin{align}
    W_{\nu e}[Y] &= (s-m_e^2)^2 G_L Y G_L + (u-m_e^2)^2 G_R Y G_R + m_e^2 t(G_L Y G_R + G_R Y G_L)\;, \notag\\ 
    W_{\nu \bar e}[Y] &= (u-m_e^2)^2 G_L Y G_L + (s-m_e^2)^2 G_R Y G_R + m_e^2 t(G_L Y G_R + G_R Y G_L)\;, \notag\\
    W_{\nu \bar\nu}[Y] &= (u-m_e^2)^2 G_L Y G_L + (t-m_e^2)^2 G_R Y G_R + m_e^2 s(G_L Y G_R + G_R Y G_L)\;.
\end{align}
The left and right coupling matrices are functions of the Weinberg angle~\citep{Weinberg:1967tq,GLASHOW1961579}, $\theta_W$, and the remaining matrix map is
\begin{equation}\label{eq:tree-level}
    G_L = \left(-\frac{1}{2} + \sin^2\theta_W \right)\mathbb{I} + \Pi_e\;, \qquad G_R = \sin^2\theta_W \mathbb{I}\;, \qquad \mathcal{T}[Y] = Y + \operatorname{Tr}[Y] \mathbb{I}\;.
\end{equation}
Our implementation instead uses the effective low-energy couplings at zero momentum transfer~\citep{Kumar:2013yoa,Erler:2013xha} following \citet{Bennett:2020zkv}, $G_L=\mathrm{diag}(g_L^e, g_L^\mu, g_L^\tau)$, $G_R=g_R\mathbb{I}$, see Table~\ref{tab:exp-inputs}. All processes have symmetry factor $S^{(r)} = 1$, with the exception of $S^{(\nu\nu;\nu\nu)} = \frac{1}{2}$. The standard collision operator of~\eqref{eq:collision-operator} is recovered when all $F$ are diagonal. We additionally follow \citet{Bennett:2020zkv} and include the momentum-independent part of the electron thermal mass in the collision operator.

\textbf{Adiabatic transfer of averaged oscillations.} 
Neutrino oscillations proceed over much shorter timescales than cosmic expansion or collisions, allowing several simplifications to the QKEs. Following~\citet{Froustey:2020mcq}, we rewrite the QKE in terms of the instantaneous (mass) eigenstates of the effective Hamiltonian as
\begin{equation}\label{eq:ATAOfull}
    Hx \frac{\partial F_m}{\partial x} = -i[\mathcal{H}_\mathrm{diag},F_m] - Hx\left[V^\dagger \frac{\partial V}{\partial x}, F_m\right]+V^\dagger \mathcal C_\nu[F] V\;,
\end{equation}
with $V$ a unitary diagonalising matrix, and where the diagonal Hamiltonian and mass-basis density matrices are \begin{equation}
    \mathcal{H}_\mathrm{eff} = V \mathcal{H}_\mathrm{diag} V^\dagger\;, \qquad F_m= V^\dagger F V\;.
\end{equation}
Adiabaticity, the slow evolution of the background, and hence the eigenbasis, allows us to neglect the second commutator in~\eqref{eq:ATAOfull}. Additionally, as many oscillations occur between successive collisions, complex phases in the off-diagonal components of $F_m$ average to zero. Consequently, we set
\begin{equation}
    F_m \to F_\mathrm{diag} = \operatorname{diag}(f_1,f_2,f_3)\;,
\end{equation} 
in terms of the instantaneous mass eigenstate occupations, $f_i$, $i \in \{1,2,3\}$.
The resulting QKE takes the form of a Boltzmann equation
\begin{equation}
    Hx\frac{\partial f_i}{\partial x}= (\mathcal{C}_{\nu,\mathrm{diag}})_i[f]\;,
\end{equation}
with the diagonal collision operator
\begin{align}
    (\mathcal{C}_{\nu,\mathrm{diag}})_i[f] &= \frac{1}{2E_1} \sum_r S^{(r)} \int \dd\Phi_2\left[G_i^{(r)}(1-f_i(y_1)) - L_i^{(r)} f_i(y_1)\right]\;,\notag\\ 
    G_i^{(r)} &= (V^\dagger \mathcal{K}_\mathcal{G}^{(r)}V)_{ii}\;, \qquad L_i^{(r)} = (V^\dagger \mathcal{K}_\mathcal{L}^{(r)}V)_{ii}\;,
\end{align}
where $\dd\Phi_2$ is the usual $2 \to 2$ invariant phase-space measure.

The transformation matrix is a function of the evolution variable $x$ and of the momentum $y$ appearing in $\mathcal{H}_\mathrm{eff}$, $V=V(x,y)$. Rather than explicitly diagonalising for each momentum value, which would be prohibitively expensive, we instead define $\mathcal{H}'_\mathrm {eff} = 2y\mathcal{H}_\mathrm{eff}$:
\begin{equation}
    \mathcal{H}'_\mathrm{eff}(x,y) = A(x) - 2y^2B(x)\;,\qquad A=U\Delta M^2U^\dagger\;,\qquad B=\frac{4G_F}{\sqrt{2}m_W^2} (\rho_e+P_e)\Pi_e\;.
\end{equation}
As rescaling a matrix does not change its eigenvectors, it suffices to find the matrix diagonalising $\mathcal{H}'_\mathrm{eff}$. To construct the transformation matrix $V(x,y)$, we first apply a $2\times 2$ block rotation, $W(x)$, to diagonalise the $\mu$-$\tau$ components, under which $\Pi_e=\mathrm{diag}(1,0,0)$ is invariant. The resulting matrix, $H(x,y)=W^\dagger(x) \mathcal{H}'_\mathrm{eff}(x,y) W(x)$, only has non-zero off-diagonal components in its electron row and column, which are related by Hermiticity across the diagonal. The eigenvalues $\lambda$, and eigenvectors $u$, of $H$ follow from solving the secular equation
\begin{equation}
    H_{ee}(y) - \lambda(y) + \frac{|H_{e\mu}|^2}{\lambda(y)-H_{\mu\mu}} + \frac{|H_{e\tau}|^2}{\lambda(y)-H_{\tau\tau}} =0\;, \qquad u_i(y) = \frac{1}{\mathcal{N}_i(y)} \begin{pmatrix}
        1 \\ \frac{H_{e\mu}^*}{\lambda_i(y)-H_{\mu\mu}} \\ \frac{H_{e\tau}^*}{\lambda_i(y)-H_{\tau\tau}}
    \end{pmatrix}\;,
\end{equation}
where all $x$ dependence is implicit and $\mathcal{N}_i$ normalises each $u_i$. The diagonalising matrix of interest is then constructed as $V(x,y) = W(x)(u_1(x,y),u_2(x,y),u_3(x,y))$.

\textbf{Flavour-diagonal limit.} Neutrino mixing is only expected to give a tiny correction to $N_\mathrm{eff}$, as coherent evolution conserves $\operatorname{Tr}[F]$, and consequently $\rho_{\nu}$ in the ultrarelativistic limit. Mixing does not therefore directly transfer energy to or from the neutrino sector, and so can only affect $N_\mathrm{eff}$ indirectly through redistributing interaction rates. Indeed,~\citet{Bennett:2020zkv} find a correction $\Delta N_{\mathrm{eff}} = 0.0005$ due to these effects. As such, the flavour-diagonal limit serves as a precision benchmark of NBEs. 

In the flavour-diagonal limit, the neutrino density matrix and effective Hamiltonian satisfy
\begin{equation}
    F_{\alpha\beta} = f_{\nu_{\alpha}} \delta_{\alpha\beta}\;, \qquad [\mathcal{H}_\mathrm{eff}, F] = 0\;.
\end{equation}
It follows that the QKE reduces to the Boltzmann equation
\begin{equation}
    H x \frac{\partial f_{\nu_\alpha}}{\partial x} = (\mathcal{C}_\nu[F])_{\alpha\alpha} \equiv \mathcal{C}_{\nu_\alpha}[f]\;.
\end{equation}
Identical initial conditions for all three neutrino flavours, assumed here, additionally imply that $f_{\nu_\mu} = f_{\nu_\tau}$. We therefore evolve two independent neutrino distributions, $f_{\nu_e}$, and $f_{\nu_\beta}$, with $\beta \in \{\mu,\tau\}$.

\textbf{Equilibrium electromagnetic bath.} Electromagnetic interactions are far more efficient than weak interactions, keeping electrons and photons in equilibrium throughout neutrino decoupling. Consequently, the evolution of the combined electromagnetic system is completely described by its temperature, $T_\mathrm{EM}$, in the absence of chemical potentials.

The equation governing the evolution of $T_\mathrm{EM}$ follows from the conservation of the stress-energy tensor, $\nabla_{\mu} T^{\mu\nu} = 0$, where $\nabla$ denotes the covariant derivative. Splitting the stress-energy tensor into its components for each sector, $T^{\mu\nu}_A$, $A \in \{\mathrm{EM},\nu\}$, the timelike component of each divergence satisfies
\begin{equation}
    \nabla_\mu T^{\mu 0}_\mathrm{EM} = Hx \left[\frac{\partial\rho_\mathrm{EM}}{\partial x} -\frac{\rho_\mathrm{EM} -3P_\mathrm{EM}}{x}\right]\;, \qquad \nabla_\mu T^{\mu0}_\nu = 2 \int \frac{\dd^3y}{(2\pi)^3} y \operatorname{Tr}\left[\operatorname{Re}\left[\mathcal{C}_\nu[F]\right]\right]\;,
\end{equation}
where the latter equation follows from $\rho_\nu = 3P_\nu$, and the combination of~\eqref{eq:QKE} and~\eqref{eq:moments}. The electromagnetic energy density therefore evolves according to
\begin{equation}\label{eq:EMevolution}
    \frac{\partial\rho_\mathrm{EM}}{\partial x} = \frac{\rho_\mathrm{EM} - 3P_\mathrm{EM}}{x} - \frac{2}{Hx}\int \frac{\dd^3y}{(2\pi)^3} y \operatorname{Tr}\left[\operatorname{Re}\left[\mathcal{C}_\nu[F]\right]\right]\;.
\end{equation}
The electromagnetic pressure and energy density are decomposed into those of the photon, electron, and finite-temperature QED corrections
\begin{equation}
    \rho_\mathrm{EM} = \rho_\gamma + \rho_e + \sum_{n >0} \rho_\mathrm{QED}^{(e^n)}\;,\qquad  P_\mathrm{EM} = P_\gamma + P_e + \sum_{n >0} P_\mathrm{QED}^{(e^n)}\;.
\end{equation}
We now define the generalised electron energy moment
\begin{equation}
    \mathcal{E}_{m,n} = \frac{2}{\pi^2}\int \dd y\,y^2 E^{m}_y\,\mathcal{F}_{e}^{(n)}\;, \qquad \mathcal{F}_{e}^{(0)} = f_e, \qquad \mathcal{F}_e^{(n+1)} = f_e(1-f_e)\frac{\partial \mathcal{F}_e^{(n)}}{\partial f_e}\;,
\end{equation}
where $E_y = E_e(y)$, such that \textit{e.g.} $\mathcal{E}_{1,0} = \rho_e$, the logarithmic moment
\begin{equation}
    \mathcal{L}_n = \int_0^\infty \dd y\int_0^\infty \dd q\, \frac{yq}{E_yE_q}\ln\left|\frac{y+q}{y-q}\right| \sum_{k=0}^{n}\binom{n}{k} E_y^k E_q^{n-k} \mathcal{F}_e^{(k)}(y) \mathcal{F}_e^{(n-k)}(q)\;,
\end{equation}
and the stress-energy tensor trace, $\Theta_e = \rho_e - 3P_e$. These satisfy the relations
\begin{equation}
    \frac{\partial\mathcal{E}_{m,n}}{\partial T_{\mathrm{EM}}} = \frac{\mathcal{E}_{m+1,n+1}}{T_\mathrm{EM}^2}\;, \qquad \frac{\partial \mathcal{L}_n}{\partial T_{\mathrm{EM}}} = \frac{\mathcal{L}_{n+1}}{T_\mathrm{EM}^2}\;, \qquad \frac{\partial\Theta_e}{\partial T_\mathrm{EM}} = \frac{m_e^2}{T_\mathrm{EM}^2} \mathcal{E}_{0,1}\;.
\end{equation}
In terms of these moments, the finite-temperature QED corrections up to $\mathcal{O}(e^3)$ are~\citep{Bennett:2019ewm}
\begin{align}
    P_\mathrm{QED}^{(e^2)} &= -\frac{e^2 T_\mathrm{EM}^2}{12 m_e^2}\Theta_e - \frac{e^2}{8m_e^4} \Theta_e^2 + \frac{e^2 m_e^2}{4\pi^4} \mathcal{L}_0\;,\notag\\ 
    P_\mathrm{QED}^{(e^3)} &= \frac{e^3}{12\pi \sqrt{T_\mathrm{EM}}} \mathcal{E}_{0,1}^{\frac{3}{2}}\;, \notag\\
    \rho^{(e^n)}_\mathrm{QED} &= T_\mathrm{EM} \frac{\partial P_\mathrm{QED}^{(e^n)}}{\partial T_\mathrm{EM}}-P_\mathrm{QED}^{(e^n)}\;.
\end{align}
At $\mathcal{O}(e^4)$ and $\mathcal{O}(e^5)$, the pressure corrections are only available for $T_\mathrm{EM}\gg m_e$, and contain a logarithm of the renormalisation scale, $\bar\mu$, and a constant~\citep{Arnold:1994eb,Zhai:1995ac}. We follow~\citet{Escudero:2025kej} and set $\bar\mu=m_e$ and relate the logarithmic terms to lower-order contributions,
\begin{align}
    P_\mathrm{QED}^{(e^4)} &= -\frac{e^2}{6\pi^2}\ln\frac{m_e}{T_\mathrm{EM}}\,\bar P_\mathrm{QED}^{(e^2)} + c_4\,e^4\,\frac{\pi^2T_\mathrm{EM}^4}{45}\;, \notag\\
    P_\mathrm{QED}^{(e^5)} &= -\frac{e^2}{4\pi^2}\ln\frac{m_e}{T_\mathrm{EM}}\,P_\mathrm{QED}^{(e^3)} + c_5\,e^5\,\frac{\pi^2T_\mathrm{EM}^4}{45}\;,
\end{align}
where $\bar P_\mathrm{QED}^{(e^2)}$ denotes $P_\mathrm{QED}^{(e^2)}$ without the $\mathcal{L}_0$ term, and $c_4\approx -1.594\cdot 10^{-3}$, $c_5\approx -6.972\cdot 10^{-4}$ are dimensionless constants. 
To solve~\eqref{eq:EMevolution}, the neural distribution function outputs $T_\mathrm{EM}$, which we use to construct $\rho_\mathrm{EM}$ and $P_\mathrm{EM}$. Network parameters are then updated using the natural gradient update through the energy conservation residual.

\section{Experiment details}
\label{app:experiment-details}

We use the same neural distribution function hyperparameters for all experiments. We use $M=50$ gamma mixture components, and use $N_\mathrm{quad}=16$ or $32$ quadrature nodes for the moment evaluation. The network that predicts the mixture parameters $\omega$ in conditional runs is a standard multilayer perceptron with 3 layers, 64 channels, and tanh activations. Our implementation uses double precision (float64) throughout.

The matrix inversion for the natural gradient update uses a Tikhonov damping of $10^{-8}$ on the diagonal, and drops eigenvalues that are lower than a factor $10^{-12}$ of the largest eigenvalue. We increase the Tikhonov damping to $10^{-5}$ when inverting the $B\times B$ Gram matrix in conditional trainings. To avoid overshooting in the natural gradient update, we introduce a learning rate $\gamma$, with default $\gamma=1$, which is decreased by a factor of $2$ until the update decreases the objective $L(\theta)$ of \eqref{eq:natgrad-setup}.

We use the Crank-Nicolson (CN) integrator scheme of \eqref{eq:schemes} to refine the solver trajectory. The resulting uncertainty estimate is used in a standard PI step size controller to determine the size of the next step, with a specified minimum and maximum step size picked separately for each experiment. The error tolerance involves a Monte Carlo uncertainty floor term in addition to the standard absolute and relative tolerances.

The fixed-point iteration of Algorithm~\ref{alg:time-step} typically converges within 10 iterations for unconditional trainings, and up to 50 iterations for conditional trainings. For conditional trainings, we specify a minimum number of iterations for each step to be conservative. For unconditional trainings, we specify a maximum budget of 30 iterations, but complete the fixed-point iteration earlier if either the residual plateaus below a certain target precision for a specified number of iterations, or if the residual is dominated by Monte Carlo noise for the same number of iterations. 

We follow the MadNIS defaults of \citet{Heimel:2023ngj} for our normalising flow, using 10 bins, and an MLP with 3 layers and 32 channels per layer. We use the Adam optimiser with learning rate $\gamma=10^{-3}$ to train the normalising flow.

\subsection{Equilibrium distributions}

\textbf{Hyperparameters.}
Neural distribution functions trained with the natural gradient method follow the description above, whereas Adam trainings use a learning rate of $\gamma=0.03$ with a cosine schedule, on the same mean-squared-error objective. We use batch size $B=300$. Training for 100 iterations takes around one second on a GPU.

\subsection{Thermalisation}

\textbf{Theory.}
The formalism of \citet{Yoon:2026rce} maps onto our expanding universe formalism with $H\to 1$, $\log x \to t$, allowing us to reuse our framework without major modifications,
\begin{equation}
    \frac{\partial f}{\partial\log x} = \frac{\mathcal{C}[f]}{H} \quad\to\quad \frac{\partial f}{\partial t} = \mathcal{C}[f]\;.
\end{equation}
We focus on their most challenging case of an $m=1$ massive particle with the processes $2\leftrightarrow 2$ and $2\leftrightarrow 3$. Both processes share the same coupling $\lambda$, such that the collision operator is directly proportional to $\lambda^2$, $\mathcal{C}[f] = \lambda^2 \mathcal{C}_0[f]$. This allows us to absorb the coupling into the time variable, $t\to t'\equiv\lambda^2 t$:
\begin{equation}
    \frac{\partial f}{\partial t} = \lambda^2\mathcal{C}_0[f] \qquad\to\qquad \mathcal{C}_0[f] = \frac{\partial f}{\lambda^2 \partial t} \equiv \frac{\partial f}{\partial t'}\;.
\end{equation}
The evolution in terms of $t'$ is therefore independent of the coupling variable $\lambda$. We use this property of the model to test the precision of our coupling-conditional training in Figure~\ref{fig:thermalisation}.

\citet{Yoon:2026rce} specifies a non-thermal initial condition, $f_\mathrm{init}(y)=(e^{(y-3)/2}+1)^{-1}$. While evolving the system via the Boltzmann equation, the coupling of different momentum modes in the collision operator pushes the species towards the Bose-Einstein distribution which is the equilibrium configuration. The energy density of the species is conserved in the process. However, the number density can be changed by the $2\leftrightarrow 3$ process. The known form of the distribution functions at $t=0$ and $t\to\infty$ combined with energy conservation allows us to compute the asymptotic number density analytically. We find $n_{t\to\infty}/n_{t=0} \approx 0.900272$, and highlight this limit in Figure~\ref{fig:thermalisation}.

We use a log-uniform prior for the condition, $p_\mathrm{prior}(\log_{10}\lambda^2) = \mathcal{U}(-0.25,0.25)$.

\textbf{Baselines.}
The semi-analytical and BEST baselines in Figure~\ref{fig:thermalisation} are evaluated using the public implementation of \citet{Yoon:2026rce}. We found that the semi-analytical result is biased at $y\to 0$ and refined their quadrature integration to reduce the shift. The tilt towards larger deviations for the NBE results at $y=10^{-1}$ is caused by this suboptimal quadrature result. BEST samples the gain and loss terms independently and uses two VEGAS iterations, so it effectively uses $4\times$ more samples per step compared to our approach for the same number of integral evaluations. The gain factors in Section~\ref{sec:thermalisation} do not include this extra factor to be conservative, and would be $4\times$ larger if the factor was included.

\textbf{Hyperparameters.}
We use batch size $B=1000$ for both pretraining and evolution, and use $N_s=1.2\cdot 10^4$ Monte Carlo samples for the evolution in Figure~\ref{fig:thermalisation}. We use $\Delta\log x_\mathrm{min}=2.5$, $\Delta\log x_\mathrm{max}=25$, and find that the code uses on average 6 fixed-point iterations per step. A full training takes around 2 hours on an A30 GPU.


\subsection{Neutrino decoupling}
\label{app:neff-experiment}

\textbf{Theory.}
We discuss the theory behind neutrino decoupling in more detail in Appendix~\ref{app:neff-theory}, and summarise the essential aspects in the following. We assume thermal equilibrium for the electromagnetic plasma, including photons, electrons, and positrons, and evolve their distribution functions following the differential equation for $\rho_\mathrm{EM}$, \eqref{eq:EMevolution}. For the three neutrino species, we instead allow a general distribution function, leading to the spectral distortions visualised in Figure~\ref{fig:neff}. If no mixing is included, we evolve the muon and tau neutrinos jointly because they interact through the same processes, and so share a distribution function in the absence of initial asymmetries. We implement neutrino oscillations using the approach of \citet{Froustey:2020mcq}.

In Table~\ref{tab:neff} we study the effect of the individual physics contributions on $N_\mathrm{eff}$ by turning on the contributions one after the other. The non-instantaneous effects on $N_\mathrm{eff}$ from the neutrino-electron interactions through the neutrino collision integral move $N_\mathrm{eff}$ away from $N_\mathrm{eff}=3$. We keep the Fermi-Dirac equilibrium shape for the second row of Table~\ref{tab:neff}, whereas the following rows use the fully flexible shape that allows them to express spectral distortions. We find that runs with Fermi-Dirac distribution shape carry a larger systematic uncertainty than the runs with fully flexible distribution shape, because they cannot express the target distribution function during training, and the way the network determines the free temperature and chemical potential depends on how the fit is performed.
The finite-temperature QED corrections contribute a $+0.01025$ correction at $\mathcal{O}(e^2)$, $-0.00095$ at $\mathcal{O}(e^3)$, and $+0.00003$ at $\mathcal{O}(e^4)+\mathcal{O}(e^5)$. We also include the electron thermal mass from there on, which contributes a $-0.00003$ correction. Finally, we include flavour oscillations following the formalism in Appendix~\ref{app:neff-theory}.
\begin{table}[tb]
    \centering
    \caption{Decomposition of $N_\mathrm{eff}$ contributions, as discussed in the text.}
    \vspace{0.2cm}
    \begin{tabular}{lll}
  \toprule
  Description & NBE result & $\Delta N_\mathrm{eff}$\\
  \midrule
  Instantaneous decoupling & $3.00000$ & --\\
  + $\nu$-$e^\pm$ interactions & $3.03414$ & $+0.03414$ \\
  + spectral distortions & $3.03407$ & $-0.00007$ \\
  + finite-temperature QED & $3.04336$ & $+0.00929$ \\
  + flavour oscillations & $3.04392$ & $+0.00056$ \\
  \bottomrule
\end{tabular}
    \label{tab:neff}
\end{table}

\textbf{Experimental inputs.}
Our $N_\mathrm{eff}$ calculation relies on the set of input parameters listed below. For the measured quantity $Q^\mathrm{phys}$ with mass dimension $d$, we use $Q=(x/m_0)^d Q^\mathrm{phys}$, $m_0=m_e^\mathrm{phys}$ to obtain the corresponding comoving quantity $Q$. In Table~\ref{tab:exp-inputs} we list all parameters where uncertainties can be neglected, following \citet{Bennett:2020zkv}.
\begin{table}[tb]
    \centering
    \caption{Experimental inputs used in the $N_\mathrm{eff}$ precision calculation.}
    \begin{tabular}{lll}
    \toprule
    Quantity & Value & Reference \\
    \midrule
    $\alpha$ & $1/137.035999084$ & \citet{Tiesinga:2021myr} \\
    $m_e^\mathrm{phys}$ & $0.51099895\,\mathrm{MeV}$ & \citet{Tiesinga:2021myr}\\
    $G_F^\mathrm{phys}$ & $1.1663787\cdot 10^{-5}\,\mathrm{GeV}^{-2}$ & \citet{ParticleDataGroup:2020ssz} \\
    $m_W^\mathrm{phys}$ & $80.379\,\mathrm{GeV}$ & \citet{ParticleDataGroup:2020ssz}\\
    $m_\mathrm{Pl}^\mathrm{phys}$ & $1.22089\cdot 10^{19}\,\mathrm{GeV}$ & \citet{ParticleDataGroup:2020ssz} \\
    $g_L^e$ & $0.727$ & \citet{Kumar:2013yoa,Erler:2013xha}\\
    $g_L^\mu = g_L^\tau$ & $-0.273$ & \citet{Kumar:2013yoa,Erler:2013xha}\\
    $g_R$ & $0.233$ & \citet{Kumar:2013yoa,Erler:2013xha}\\
    \bottomrule
    \end{tabular}
    \label{tab:exp-inputs}
\end{table}

The neutrino sector parameters with $1\sigma$ uncertainties are taken from the more recent NuFIT-6.1 fit~\citep{Esteban:2024eli,NuFIT61}, and we assume $\delta_\mathrm{CP}=0$ in the PMNS matrix,
\begin{align}
    &\sin^2\theta_{12} = 0.3088_{-0.0066}^{+0.0067}\;,\quad 
    \Delta m_{21}^{2,\mathrm{phys}} = 7.537_{-0.10}^{+0.094}\cdot 10^{-5}\,\mathrm{eV}^2\;,\notag\\
    \mathrm{NO}:\quad&\sin^2\theta_{13} = 0.02249\pm 0.00057\;,\quad 
    \sin^2\theta_{23} = 0.470_{-0.014}^{+0.017}\;,\notag\\
    &\Delta m_{31}^{2,\mathrm{phys}} = 2.521_{-0.018}^{+0.026}\cdot 10^{-3}\,\mathrm{eV}^2\;,\notag\\
    \mathrm{IO}:\quad&\sin^2\theta_{13} = 0.02261\pm 0.00056\;,\quad 
    \sin^2\theta_{23} = 0.555_{-0.016}^{+0.013}\;,\notag\\
    &\Delta m_{32}^{2,\mathrm{phys}} = -2.500_{-0.023}^{+0.024}\cdot 10^{-3}\,\mathrm{eV}^2\;.
\end{align}

\textbf{Hyperparameters.}
We use a batch size $B=4000$, and $N_s=4000$ Monte Carlo samples for the collision integral. We evolve the system from $x = 10^{-2}$ to $\log x=5.0$. We use a minimum step size of $\Delta \log x=0.0025$, a maximum step size of $\Delta \log x=0.025$, alongside relative and absolute tolerances of $\epsilon_\mathrm{rel}=10^{-6}$ and $\epsilon_\mathrm{abs}=10^{-8}$, respectively, for the step size controller. For each training, we iterate on the fixed-point objective until the objective improves by less than $r_\mathrm{plateau}=0.05$ over 3 iterations, and terminate after $30$ iterations at most. We use 64 quadrature nodes for the electron moment integrals and 32 quadrature nodes for the neutrinos. We do not use the MadNIS neural importance sampling refinement for this calculation. A single training with these settings takes around 4 hours on an NVIDIA A30 GPU.

\textbf{Solver uncertainty.}
Here we describe how we have quantified the solver uncertainty, which is significantly smaller for NBEs as compared to existing approaches, and as such a key result of our work. We first performed $n=20$ trainings on independent seeds with the default hyperparameters to extract the mean, $\mu_0$, and per-run standard deviation, $\sigma_0$, reference values for $N_\mathrm{eff}$:
\begin{equation}
    \mu_0 = 3.0439206\;,\qquad \sigma_0 = 1.3\cdot 10^{-6}\;.
\end{equation}
Any digits beyond these varied with different seeds or hyperparameters, and so we do not report them in our main result, \eqref{eq:neff-result}. We next varied each of the hyperparameters above to increase precision, and estimated uncertainties based on the mean $\mu_{\delta,i}$ and per-run standard deviation $\sigma_{\delta,i}$ from $m=6$ runs. Based on these values, we extract the statistical uncertainty $\sigma_\mathrm{stat}$ as the standard deviation of the mean $\mu_0$ and the systematic uncertainties $\sigma_{\mathrm{syst},i}$ from varying each of the hyperparameters,
\begin{equation}
    \sigma_\mathrm{stat} = \frac{\sigma_0}{\sqrt{n}}\;,\qquad \sigma_{\mathrm{syst},i} = \max \left(|\mu_0 - \mu_{\delta,i}|, \sigma_{\delta,i}\sqrt{1/n+1/m}\right)\;.
\end{equation}
The systematic effects are summarised in Table~\ref{tab:neff-uncertainty-solver}. For Monte Carlo sampling, we vary the batch size $4000\to 8000$ and the number of collision integral samples $4000\to 8000$. We increase the number of quadrature nodes by a factor of $2$. We start the evolution at $x=10^{-2}\to 10^{-3}$. For the step size controller, we decrease the minimum step size $\Delta\log x_\mathrm{min}=0.0025\to 0.000625$, the maximum step size $\Delta\log x_\mathrm{max}=0.025\to 0.00625$, and the tolerances $\epsilon_\mathrm{rel}=10^{-6}\to 4\cdot 10^{-7}$, $\epsilon_\mathrm{abs}=10^{-8}\to 2.5\cdot 10^{-9}$. For the fixed-point iteration convergence criterion, we increase the maximum number of iterations $30\to 60$ and decrease the early exit plateau $r_\mathrm{plateau}=0.05\to 0.0125$. All hyperparameters are varied independently and added in quadrature, $\sigma_\mathrm{tot} = (\sum_i \sigma_i^2)^{1/2}$, which corresponds to the assumption of independent Gaussian uncertainties. For the step size controller and fixed-point iteration convergence values in Table~\ref{tab:neff-uncertainty-solver} we already combined several individual effects. Finally, we combine all statistical and systematic effects, and obtain a total solver uncertainty of $2.0\cdot 10^{-6}$. To further validate this procedure, we repeat the procedure for a lower-precision calculation in Appendix~\ref{app:neff-uncertainty-validation} and find that this uncertainty estimate covers the result of the higher-precision calculation performed here.

\begin{table}[tb]
    \centering
    \caption{Contributions to the $\pm 2\cdot 10^{-6}$ solver uncertainty in the $N_\mathrm{eff}$ calculation in \eqref{eq:neff-result}. We quantify the effects by increasing the precision for every parameter individually, and combine the resulting uncertainties as independent Gaussian effects.}
    \label{tab:neff-uncertainty-solver}
    \begin{tabular}{l r}
    \toprule
    Effect & $10^6\cdot \delta N_\mathrm{eff}$ \\
    \midrule
    Seed statistics & $\pm 0.3$ \\
    \midrule
    Monte Carlo sampling (batch size) & $\pm 0.6$ \\
    Monte Carlo sampling (collision integral) & $\pm 0.6$ \\
    Moment quadrature & $\pm 0.7$ \\
    Evolution starting point & $\pm 0.6$ \\
    Step size controller & $\pm 1.4$ \\
    Fixed-point iteration convergence & $\pm 0.6$ \\
    \midrule
    Total & $\pm 2.0$ \\
    \bottomrule
    \end{tabular}
\end{table}

\textbf{Experimental uncertainty.}
We now study how uncertainties from the NuFIT-6.1 global fits on neutrino mixing angles $\sin^2\theta_{ij}$ and mass differences $\Delta m_{ij}^2$ propagate into uncertainties on the $N_\mathrm{eff}$ determination. To this end, we follow \citet{Bennett:2020zkv} and run solvers with input parameters varied by $3\sigma$ upwards and downwards. We then average the two resulting shifts in $N_\mathrm{eff}$ to obtain an uncertainty estimate. We also perform one run with inverted neutrino mass ordering, finding a one-sided shift. Table~\ref{tab:neff-uncertainty-experiment} shows the resulting uncertainties, which are dominated by the solar mixing angle $\theta_{12}$. This is in agreement with \citet{Bennett:2020zkv}, where the authors found that the dominating uncertainty comes from the solar mixing angle, although as a $10^{-4}$ effect due to the larger experimental uncertainty on $\theta_{12}$ when the study was performed. 

\begin{table}[tb]
    \centering
    \caption{Contributions to the $\pm 2\cdot 10^{-5}$ experimental uncertainty in the $N_\mathrm{eff}$ calculation in \eqref{eq:neff-result}. We quantify the effects by varying the input parameters within their reported experimental uncertainty band, and for the neutrino mass ordering we report the effect of flipping the choice.}
    \label{tab:neff-uncertainty-experiment}
    \begin{tabular}{l r}
    \toprule
    Effect & $10^6\cdot \delta N_\mathrm{eff}$ \\
    \midrule
    $\sin^2\theta_{12}$ ($3\sigma$ band) & $\pm 18$ \\
    $\sin^2\theta_{13}$ ($3\sigma$ band) & $\pm 4$ \\
    $\Delta m^2_{21}$ ($3\sigma$ band) & $\pm 3$ \\
    Neutrino mass ordering (NO $\to$ IO) & $+7$ \\
    \midrule
    Total & $\pm 20$ \\
    \bottomrule
    \end{tabular}
\end{table}

\textbf{Theory uncertainty.}
On the theory side, there are several effects that are expected to yield contributions at the level of experimental uncertainty, $2\cdot 10^{-5}$, or above. Quantifying this uncertainty is challenging, as it ultimately requires the full calculation. A first estimate can be obtained from approximate calculations, however we do not include these estimated effects as uncertainties in the same way as the solver or experimental uncertainties as they strongly depend on the quality of these approximations, and instead discuss them here. Most importantly, QED corrections to the neutrino collision integral are expected to contribute at the $10^{-4}$ level, based on approximations ranging from $7\cdot 10^{-4}$~\citep{Cielo:2023bqp} to $2\cdot 10^{-5}$~\citep{Drewes:2024wbw}. \citet{Binder:2024vmy} estimate the impact of positronium formation and find corrections to $N_\mathrm{eff}$ at the level of $10^{-4}$ to $10^{-6}$ depending on the prescription. Finite-temperature corrections to the QED equation of state are known up to $\mathcal{O}(e^3)$ including electron mass effects, and up to $\mathcal{O}(e^5)$ in the $T\gg m_e$ limit, see Appendix~\ref{app:neff-theory}. Finally, our approximation of adiabatic neutrino oscillations in Appendix~\ref{app:neff-theory} is estimated to affect the value of $N_\mathrm{eff}$ below the $10^{-6}$ level. We emphasise that the significantly reduced solver uncertainty of the NBE approach opens the door for probing many of these precision effects.

\textbf{Comparison with literature results.}
We now discuss the differences between our approach and the implementation of \citet{Bennett:2020zkv}, which we used as a reference. First, our flavour oscillation treatment follows \citet{Froustey:2020mcq}, see Appendix~\ref{app:neff-theory}, which induces a deviation at the $10^{-6}$ level or below. Second, we include finite-temperature QED corrections up to $\mathcal{O}(e^5)$, instead of only $\mathcal{O}(e^3)$, see Appendix~\ref{app:neff-theory}. Third, we use the recent NuFIT-6.1 values for the neutrino masses and mixing angles, with significantly reduced uncertainties compared to the values used by \citet{Bennett:2020zkv}, translating into reduced uncertainties in $N_\mathrm{eff}$. As a check, we repeat the $N_\mathrm{eff}$ calculation following the \citet{Bennett:2020zkv} choices in all the cases above except the flavour oscillations, finding
\begin{equation}
    N_\mathrm{eff} = 3.04390\;\pm 2\cdot 10^{-6}\,(\text{solver})\;\pm 4\cdot 10^{-5}\,(\text{experiment})\;.
\end{equation}
This is in agreement with the $N_\mathrm{eff}=3.0440\pm 0.0001(\mathrm{solver})\pm 0.0001(\mathrm{experiment})$ result of \citet{Bennett:2020zkv}. However, we find a smaller experimental uncertainty from varying $\sin^2\theta_{12}=0.318\pm 0.048$ on a $3\sigma$ band.

\textbf{New physics parameter scan.}
The new physics parameter scan in Figure~\ref{fig:neff} does not require the $10^{-6}$ precision of the calculation described above, here $10^{-2}$ precision is sufficient to understand whether the new theory with dark photon $A'$ and complex scalar $\phi$ can shift the value of $N_\mathrm{eff}$ by the correct amount. We therefore use the `fast' hyperparameters of Appendix~\ref{app:neff-uncertainty-validation} for this study, see Table~\ref{tab:neff-solver-settings}. We use a uniform prior for $r\in [2.01,2.7]$ and a log-uniform prior for $m_\phi\in [1,20]\,\mathrm{MeV}$. The conditional training requires an accurate evolution of the neural distribution function over the full prior range, and we therefore additionally increase the batch size to 1000. We also require at least 15 iterations per step to ensure that the prior space is sufficiently explored. A single run with these settings takes around 5 hours on an NVIDIA A30 GPU.

We validated the results of the conditional run by comparing with unconditional runs using the `fast' hyperparameters of Table~\ref{tab:neff-solver-settings}, which are expected to have a solver uncertainty of $\delta N_\mathrm{eff}\approx 10^{-5}$. The $N_\mathrm{eff}$ results of the conditional training vary from this reference by $10^{-3}$ to $10^{-4}$ depending on the theory parameters, which is sufficient for phenomenological studies.

\section{Additional results}
\label{app:additional-results}



\subsection{Number of mixture components in neural distribution functions}

\begin{figure}[b!]
    \centering
    \includegraphics[width=0.5\linewidth]{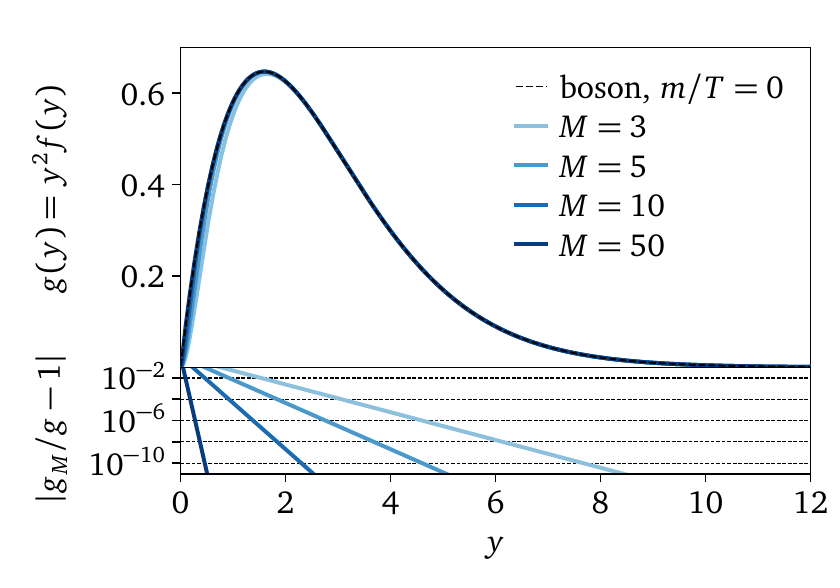}
    \vspace{-0.6cm}
    \caption{Convergence speed of the gamma mixture series for massless bosons. We compare the exact Bose-Einstein distribution with the shapes obtained after truncating the series after $M$ terms.}
    \label{fig:boson_ndist}
\end{figure}

Fermionic and bosonic equilibrium distribution functions for massless particles can be exactly represented using the gamma mixture series in Section~\ref{sec:neural_distribution_functions}, using a single mixture component for fermions and an infinite number of mixture components for bosons, see Appendix~\ref{app:distribution-function}. We visualise the bosonic case in Figure~\ref{fig:boson_ndist}, where we show the exact Bose-Einstein shape compared with the shapes obtained by truncating the gamma mixture series after $M$ terms. The series converges slowly for small $y$ values due to the $e^{-k\varepsilon}$ term in the series representation, \eqref{eq:feq-series}.

\begin{figure}[b!]
    \centering
    \includegraphics[width=0.49\linewidth,page=1]{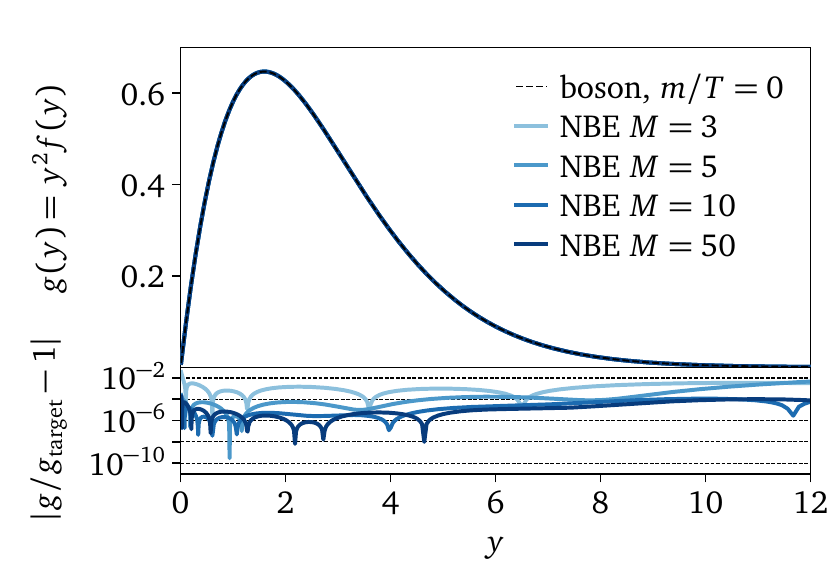}
    \includegraphics[width=0.49\linewidth,page=2]{figs/equilibrium_ndist.pdf}
    \includegraphics[width=0.49\linewidth,page=3]{figs/equilibrium_ndist.pdf}
    \includegraphics[width=0.49\linewidth,page=4]{figs/equilibrium_ndist.pdf}
    \vspace{-0.6cm}
    \caption{Trained neural distribution functions with different numbers of mixture components $M$, for massless and massive fermions and bosons.}
    \label{fig:equilibrium_ndist}
\end{figure}

In Figure~\ref{fig:equilibrium_ndist} we show the neural distribution functions learned with NBE, for the four cases of fermionic/bosonic and massless/massive particles in equilibrium, extending the left panel of Figure~\ref{fig:equilibrium}. We follow the same training as in Section~\ref{sec:detailed balance}, described in Appendix~\ref{app:experiment-details}. We vary the number of mixture components between $M=3$ and our default $M=50$, and compare to the truth distribution. We find that $M=3$ already achieves $10^{-4}$ precision, and the precision plateaus around $M=10$, with no significant improvements when going to our default $M=50$ in this case. For massless fermions which can be expressed with a single mixture component, we instead find that $M=3$ achieves the highest precision at $10^{-10}$, and more mixture components degrade to $10^{-6}$ due to the noise in the higher-dimensional matrix inversion required for the natural gradient step. Compared to Figure~\ref{fig:boson_ndist}, we see that neural distribution functions do not learn the gamma mixture series representation of the distribution function which converges slowly for small $y$, but a representation with the highest precision at intermediate $y$ values, which receive the largest weight in the objective. For $M=3$, the learned representation in Figure~\ref{fig:equilibrium_ndist} is on average more precise than the series representation in Figure~\ref{fig:boson_ndist}.

\subsection{Longer Adam training for equilibrium distributions}

\begin{figure}
    \centering
    \includegraphics[width=0.49\linewidth,page=3]{figs/equilibrium.pdf}
    \includegraphics[width=0.49\linewidth]{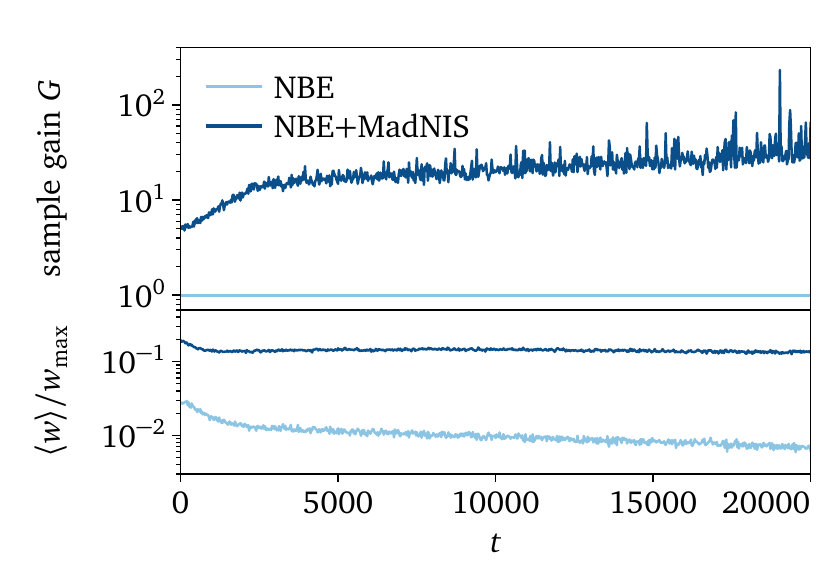}
    \vspace{-0.6cm}
    \caption{Left: Longer Adam training for massless equilibrium distribution functions. Right: Benefit of neural importance sampling for the thermalisation experiment in Section~\ref{sec:thermalisation}, comparing NBE with and without the neural importance sampling refinement using MadNIS.}
    \label{fig:best_madnis}
\end{figure}

Section~\ref{sec:detailed balance} discusses the benefits of natural gradient training for neural distribution functions compared to training with the Adam optimiser. We find in Figure~\ref{fig:equilibrium} that Adam training converges slowly, whereas natural gradient training reaches a plateau at $10^{-6}$ deviation within 10 iterations. To understand the limitations of the Adam optimiser for this problem, we perform a longer training with 10,000 instead of 100 iterations, see the left part of Figure~\ref{fig:best_madnis}. We find that Adam then reaches $10^{-4}$ deviation, still significantly above the value reached by natural gradient after only 10 iterations.

\subsection{Gain from neural importance sampling}

Neural importance sampling can be used to refine the Monte Carlo samples in Algorithm~\ref{alg:collision-sampling}. We study the gain from neural importance sampling for the thermalisation experiment, training on a single coupling and otherwise using the same hyperparameters as Section~\ref{sec:thermalisation}, described in Appendix~\ref{app:experiment-details}.

We study the gain of neural importance sampling over the thermalisation process for the full $\phi\phi\leftrightarrow\phi\phi$ and $\phi\phi\leftrightarrow\phi\phi\phi$ interactions in the right part of Figure~\ref{fig:best_madnis}. In the top panel, we show the effective sample gain compared to the NBE baseline obtained by requiring $\sigma/\sqrt{N} = \sigma_\mathrm{NBE}/\sqrt{N_\mathrm{NBE}}$, to get effective sample gain $G=N_\mathrm{NBE}/N$. Whereas we find only $G\approx 5$ at $t=0$, the gain $G$ increases to $G\approx 20$ while the system thermalises. This is due to the change of the integrand and neural distribution function density, which modify the gap between the default and optimal proposal density that can be closed by neural importance sampling. In the lower panel of Figure~\ref{fig:best_madnis}, we study the unweighting efficiency $\langle w\rangle / w_\mathrm{max}$ as a direct measure of the distribution of integral weights $w$. We see the same $10\times$ increase in sample efficiency as in the first panel.

\subsection{Natural gradient update with conserved moments}
\label{app:moment-pins}

\begin{figure}
    \centering
    \includegraphics[width=0.49\linewidth,page=1]{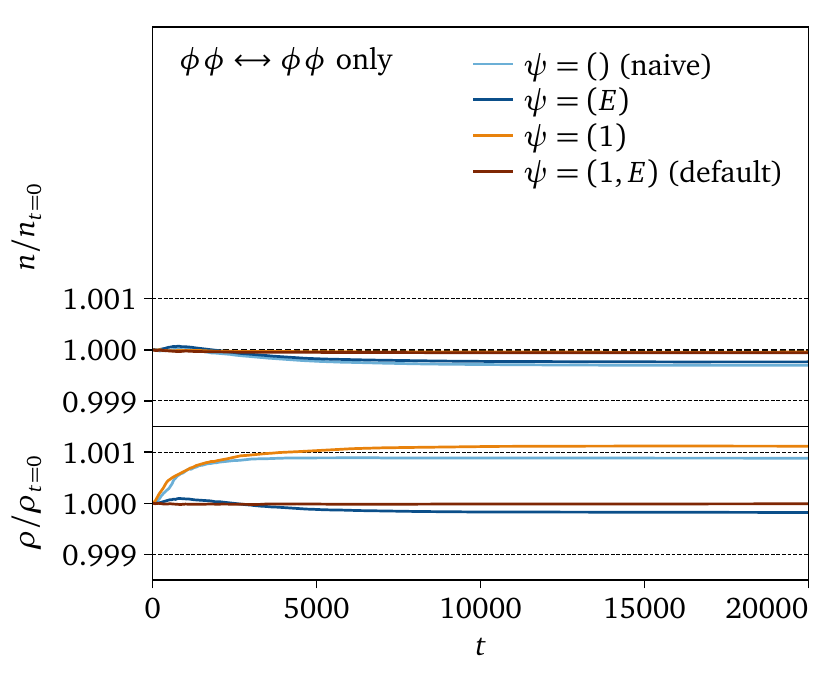}
    \includegraphics[width=0.49\linewidth,page=2]{figs/best_pins.pdf}
    \vspace{-0.6cm}
    \caption{Effect of the moment conservation trick of Appendix~\ref{app:training} for scalar thermalisation, including only $2\leftrightarrow 2$ processes (left) and also $2\leftrightarrow 3$ processes (right).}
    \label{fig:best_pins}
\end{figure}

In Appendix~\ref{app:training} we discussed a modification tothe natural gradient update that improves the handling of conserved moments. We test the effect of this trick by evolving the thermalisation experiment of Section~\ref{sec:thermalisation} with different choices of conserved moments $\psi=(\psi_1,\dots \psi_n)$, see Appendix~\ref{app:training} for the formalism. The naive baseline $\psi =()$ does not use the trick, and the extensions $\psi=(E)$, $\psi=(1)$, $\psi =(1,E)$ improve the handling of energy density conservation, number density conservation, and both. The left panel of Figure~\ref{fig:best_pins} includes only the $\phi\phi\leftrightarrow\phi\phi$ process, such that both energy and number density are conserved, whereas the right panel includes also $\phi\phi\leftrightarrow\phi\phi\phi$ processes such that number density is no longer conserved. For both systems, we find that the $\psi=(1,E)$ run achieves the highest precision, with $10^{-4}$ or lower relative deviation, whereas the naive approach only achieves $10^{-3}$ deviation.

\subsection{Fast neutrino decoupling solver}
\label{app:neff-uncertainty-validation}

\begin{table}[tb]
    \centering
    \begin{small}
    \caption{Solver settings varied for the uncertainty estimate of Table~\ref{tab:neff-uncertainty-solver2}. We write $a\to b$ for each entry, with $a$ the default value and $b$ the increased-precision value used for the solver uncertainty estimate.}
    \label{tab:neff-solver-settings}
    \begin{tabular}{lll}
    \toprule
    Setting & precision & fast \\
    \midrule
    Batch size & $4000 \to 8000$ & $500 \to 1000$ \\
    Collision integral samples & $4000 \to 8000$ & $4000 \to 8000$ \\
    \midrule
    Neutrino moment quadrature & $32 \to 64$ & same \\
    EM-sector moment quadrature & $64 \to 128$ & same \\
    Photon series truncation & $1000 \to 2000$ & same \\
    QED two-dimensional quadrature & $32 \to 64$ & same \\
    \midrule
    $x_\mathrm{min}$ & $10^{-2} \to 10^{-3}$ & same \\
    \midrule
    $\Delta \log x_\mathrm{max}$ & $0.025 \to 0.00625$ & same \\
    $\Delta \log x_\mathrm{min}$ & $0.0025 \to 6.25\cdot 10^{-4}$ & $0.005 \to 1.25\cdot 10^{-3}$ \\
    Relative tolerance $\epsilon_\mathrm{rel}$ & $10^{-6} \to 4\cdot 10^{-7}$ & $6.25\cdot 10^{-6} \to 1.5625\cdot 10^{-6}$ \\
    Absolute tolerance $\epsilon_\mathrm{abs}$ & $10^{-8} \to 2.5\cdot 10^{-9}$ & same \\
    \midrule
    Fixed-point plateau threshold $r_\mathrm{plateau}$ & $0.05 \to 0.0125$ & same \\
    Fixed-point maximum iterations & $30 \to 60$ & same \\
    \bottomrule
    \end{tabular}
    \end{small}
\end{table}

\begin{table}[tb]
    \centering
    \caption{Contributions to the solver uncertainty in the $N_\mathrm{eff}$ calculation, for the two
    solver settings of Table~\ref{tab:neff-solver-settings}. We quantify the effects by increasing the
    precision for every parameter individually and combine the resulting uncertainties as independent
    Gaussian effects.}
    \label{tab:neff-uncertainty-solver2}
    \begin{tabular}{lrr}
    \toprule
    Effect & $10^6\cdot \delta N_\mathrm{eff}$ (precision) & $10^6\cdot \delta N_\mathrm{eff}$ (fast) \\
    \midrule
    Seed statistics & $\pm 0.3$ & $\pm 1.1$ \\
    \midrule
    Monte Carlo sampling (batch size) & $\pm 0.6$ & $\pm 3.4$ \\
    Monte Carlo sampling (collision integral) & $\pm 0.6$ & $\pm 2.4$ \\
    Moment quadrature & $\pm 0.7$ & $\pm 4.9$ \\
    Evolution starting point & $\pm 0.6$ & $\pm 2.4$ \\
    Step size controller & $\pm 1.4$ & $\pm 6.5$ \\
    Fixed-point iteration convergence & $\pm 0.6$ & $\pm 2.4$ \\
    \midrule
    Total & $\pm 2.0$ & $\pm 9.8$ \\
    \bottomrule
    \end{tabular}
\end{table}

To validate the solver uncertainty quantification procedure used in Appendix~\ref{app:neff-experiment}, we repeat the procedure for a calculation with hyperparameters that achieve reduced precision. We can then use the result of the precision calculation to validate the uncertainty estimate of the lower-precision calculations.

We specify the default and varied parameters in Table~\ref{tab:neff-solver-settings}, with `precision' for our defaults used in Appendix~\ref{app:neff-experiment}, and `fast' for the additional study in this section. A `precision' run takes around 4 hours on an NVIDIA A30 GPU, whereas a `fast' run only takes 40 minutes. Both runs use the experimental uncertainty and the same assumptions from the theory side. We follow the procedure of Appendix~\ref{app:neff-experiment} to obtain the solver uncertainty estimates, and find the results in Table~\ref{tab:neff-uncertainty-solver2}. In total, we find
\begin{align}
    N_\mathrm{eff} &= 3.043915 \;\pm 0.000010\,(\mathrm{solver})\quad \text{(fast)}\;,\notag\\
    N_\mathrm{eff} &= 3.043921 \;\pm 0.000002\,(\mathrm{solver})\quad \text{(precision)}\;.
\end{align}
The solver uncertainty of the fast run covers the result of the more precise calculation, validating the uncertainty estimate procedure of Appendix~\ref{app:neff-experiment}.